\documentclass[12pt]{article}

\usepackage[margin=1.0in]{geometry}

\usepackage{booktabs}      
\usepackage{siunitx}       
\usepackage{multirow}      
\usepackage{tabularx}      
\usepackage{makecell}      
\usepackage{geometry}      
\usepackage{cite}

\usepackage{amssymb,latexsym}

\usepackage{epstopdf}

\usepackage{graphics,epsfig}

\usepackage{color}
\usepackage{xcolor}

\usepackage{amsmath,amsfonts}

\usepackage{mathtools}

\usepackage{mathrsfs}

\usepackage{slashed}

\usepackage{hyperref}

\usepackage{array}

\DeclareMathAlphabet{\mathpzc}{OT1}{pzc}{m}{it}

\usepackage{float}

\usepackage{feynmp}
\usepackage{feynmp-auto}

\let\a=\alpha \let\b=\beta \let\g=\gamma \let\d=\delta \let\e=\epsilon
\let\z=\zeta  \let\th=\theta  \let\k=\kappa
\let\l=\lambda \let\m=\mu \let\n=\nu \let\x=\xi \let\p=\pi 
\let\s=\sigma \let\t=\tau \let\u=\upsilon \let\f=\phi  
 
\let\w=\omega       \let\D=\Delta \let\Th=\Theta \let\L=\Lambda
\let\X=\Xi  \let\S=\Sigma  \let\Y=\Psi
 
\let\la=\label  
  
\def\nn{\nonumber} \def\bd{\begin{document}} \def\ed{\end{document}}
\let\fr=\frac \let\bl=\bigl \let\br=\bigr
\let\Br=\Bigr \let\Bl=\Bigl
\let\bm=\bibitem
\let\na=\nabla
\def\tU{{\widetilde U}}
\let\pa=\partial \let\ov=\overline
\def\ie{{\it i.e.\ }}
\newcommand{\be}{\begin{equation}}
\newcommand{\ee}{\end{equation}}
\def\ba{\begin{array}}
\def\ea{\end{array}}
\def\ft#1#2{{\textstyle{{\scriptstyle #1}\over {\scriptstyle #2}}}}
\def\fft#1#2{{#1 \over #2}}
\def\F#1#2{{ F_{#1}^{(#2)} }}
\def\cF#1#2{{ {\cal F}_{#1}^{(#2)} }}

\def\R{{\bf R}}
\def\sst#1{{\scriptscriptstyle #1}}
\def\oneone{\rlap 1\mkern4mu{\rm l}}
\def\e7{E_{7(+7)}}
\def\td{\tilde}
\def\wtd{\widetilde}
\def\im{{\rm i}}
\def\bog{Bogomol'nyi\ }
\newcommand{\ho}[1]{$\, ^{#1}$}
\newcommand{\hoch}[1]{$\, ^{#1}$}
\newcommand{\bea}{\begin{eqnarray}}
\newcommand{\eea}{\end{eqnarray}}
\newcommand{\ra}{\rightarrow}
\newcommand{\lra}{\longrightarrow}
\newcommand{\Lra}{\Leftrightarrow}
\newcommand{\ap}{\alpha^\prime}
\newcommand{\bp}{\tilde \beta^\prime}
\newcommand{\cB}{{\cal B}}
\newcommand{\cO}{{\cal O}}
\newcommand{\vecx}{\vec{x}}
\newcommand{\vecy}{\vec{y}}
\newcommand{\vecp}{\vec{p}}
\newcommand{\vecq}{\vec{q}}
\newcommand{\tr}{{\rm tr} }
\newcommand{\Tr}{{\rm Tr} }
\newcommand{\NP}{Nucl. Phys. }

\newcommand{\cL}{{\cal L}}
\newcommand{\cA}{{\cal A}}
\newcommand{\cT}{{\cal T}}
\newcommand{\cR}{{\cal R}}
\newcommand{\cD}{{\cal D}}
\newcommand{\cH}{{\cal H}}

\def\Cb{\bar{C}}

\def\sst#1{{\scriptscriptstyle #1}}
\def\0{{\sst{(0)}}}
\def\1{{\sst{(1)}}}
\def\2{{\sst{(2)}}}
\def\3{{\sst{(3)}}}
\def\4{{\sst{(4)}}}
\def\5{{\sst{(5)}}}
\def\6{{\sst{(6)}}}
\def\7{{\sst{(7)}}}
\def\8{{\sst{(8)}}}
\def\9{{\sst{(9)}}}
\def\p{{\sst{(p)}}}
\def\q{{\sst{(q)}}}
\def\ve{\varepsilon}
\def\vf{\varphi}
\def\F{\Phi}
\def\wg{\wedge}

\def\e{\epsilon}
\def\barl{\bar{l}}

\def \bi{\bibitem}
\def \la {\label}

\def \l {\lambda}
\def\foot{\footnote}
\def \tl  {{\tilde \l}}
\def \sql {{\sqrt \l}}
\def \adss {$AdS_5 \times S^5$\ }
\newcommand{\rf}[1]{(\ref{#1})}
\def \ov {\over}

\def\th{\theta}
\def\Th{\Theta}
\def\vth{\vartheta}
\def\btheta{{\bar\theta}}
\def\ttheta{{{\tilde\theta}}}
\def\bttheta{{{\bar\ttheta}}}
\def\vth{\vartheta}

\def\ra{\rightarrow}
\def\N{\nabla}
\def\F{{\cal F}}
\def\uM{\underline{M}}
\def\uA{\underline{A}}
\def\uN{\underline{N}}
\def\uP{\underline{P}}
\def\ua{\underline{a}}
\def\ub{\underline{b}}
\def\uc{\underline{c}}
\def\ud{\underline{d}}
\def\ue{\underline{e}}
\def\uf{\underline{f}}
\def\ui{\underline{i}}
\def\uj{\underline{j}}
\def\uk{\underline{k}}
\def\ul{\underline{l}}
\def\ual{\underline{\alpha}}
\def\ube{\underline{\beta}}
\def\um{\underline{m}}
\def\un{\underline{n}}
\def\up{\underline{p}}
\def\uq{\underline{q}}
\def\ur{\underline{r}}
\def\us{\underline{s}}
\def\umu{\underline{\mu}}
\def\unu{\underline{\nu}}
\def\ula{\underline{\l}}
\def\uka{\underline{\k}}
\def\usi{\underline{\s}}
\def\urh{\underline{\r}}
\def\cc{\circ}
\def\eqv{\equiv}

\def\ni{\noindent}

\def\Ep{E^{{}^{(+)}}}
\def\Em{E^{{}^{(-)}}}

\def\Mp{M^{{}^{(+)}}}
\def\Mm{M^{{}^{(-)}}}

\def \ha{{1\ov 2}}

\def\r{\rho}

\def\Y{{\rm Y}}
\def\X{{\rm X}}
\def\tY{\tilde{\rm Y}}
\def\tX{\tilde{\rm X}}
\def\dY{\dot{\rm Y}}
\def\dX{\dot{\rm X}}

\def \J {\mathcal{J}}
\def \del {\partial}

\def\dF{\dot{F}}
\def\dG{\dot{G}}
\def\df{\dot{f}}
\def \E {{\cal E}}
\def \S {{\cal S}}
\def \J {{\cal J}}

\def\ms{\mathcal{S}}
\def\mj{\mathcal{J}}
\def\soj{\fr{\ms}{\mj}}
\def \R {{\bf R}}
\def \om {\omega}
\def \bE {\bar E}
\def \x {{\cal X}}

\def \bi{\bibitem}
\def \la {\label}

\def \l {\lambda}
\def\foot{\footnote}
\def \tl  {{\tilde \l}}
\def \sql {{\sqrt \l}}
\def \adss {$AdS_5 \times S^5$\ }
\def \ov {\over}

\def \varpi {{\rm w}}

\def\thb{\bar{\theta}}
\def\Thb{\bar{\Theta}}
\def\barp{\bar{p}}
\def\barq{\bar{q}}
\def\barc{\bar{c}}
\def\bard{\bar{d}}
\def\bare{\bar{e}}

\def\thb{\bar{\theta}}
\def\Thb{\bar{\Theta}}
\def\mb{\bar{\m}}
\def\ab{\bar{\a}}
\def\zb{\bar{z}}
\def\psib{\bar{\psi}}
\def\barl{\bar{l}}
\def\barp{\bar{p}}
\def\barq{\bar{q}}
\def\barc{\bar{c}}
\def\bard{\bar{d}}
\def\baru{\bar{u}}

\def\e{\epsilon}
\def\wb{\bar{w}}
\def\lb{\bar{\l}}
\def\Jb{\bar{J}}
\def\Nb{\bar{N}}
\def\Zb{\bar{Z}}
\def\pab{\bar{\pa}}

\def\At{\tilde{A}}
\def\Bt{\tilde{B}}
\def\Ct{\tilde{C}}
\def\Dt{\tilde{D}}
\def\Et{\tilde{E}}
\def\Ft{\tilde{F}}
\def\Gt{\tilde{G}}
\def\Ht{\tilde{H}}
\def\Kt{\tilde{K}}
\def\Mt{\tilde{M}}
\def\Nt{\tilde{N}}
\def\Rt{\tilde{R}}
\def\at{\tilde{a}}
\def\bt{\tilde{b}}
\def\ct{\tilde{c}}
\def\dt{\tilde{d}}
\def\et{\tilde{e}}
\def\ft{\tilde{f}}
\def \zt{\tilde{z}}
\def \ztt{\tilde{\z}}
\def\Omt{\tilde{\Omega}}
\def \zetat{\tilde{\zeta}}
\def\htil{\tilde{h}}
\def\gt{\tilde{g}}
\def\nt{\tilde{n}}
\def\mut{\tilde{\mu}}
\def\nut{\tilde{\nu}}
\def\pht{\tilde{\f}}
\def\Phit{\tilde{\Phi}}
\def\vft{\tilde{\vf}}
\def\etat{\tilde{\eta}}

\def\rht{\tilde{\rho}}

\def\asth{\hat{*}}
\def\phh{\hat{\phi}}

\def\bA{{\bf A}}

\def\ola{\overleftarrow}
\def\ora{\overrightarrow}
\def\alt{\tilde{\a}}

\def\eh{\hat{e}}
\def\eph{\hat{\e}}
\def\ph{\hat{p}}
\def\alh{\hat{\a}}
\def\beh{\hat{\b}}
\def\gah{\hat{\g}}
\def\Fh{\hat{F}}
\def\muh{\hat{\m}}
\def\nuh{\hat{\n}}
\def\thh{\hat{\th}}
\def\rhh{\hat{\r}}
\def\dh{\hat{d}}
\def\ih{\hat{i}}
\def\jh{\hat{j}}
\def\hh{\hat{h}}
\def\nh{\hat{n}}
\def\gh{\hat{g}}
\def\kh{\hat{k}}
\def\deh{\hat{\d}}
\def\wh{\hat{w}}
\def\lah{\hat{\l}}
\def\Ah{\hat{A}}
\def\Gh{\hat{G}}
\def\Kh{\hat{K}}
\def\Nh{\hat{N}}
\def\Rh{\hat{R}}
\def\Ch{\hat{C}}
\def\Omh{\hat{\Omega}}

\def\xh{\hat{x}}

\def\ps{\rlap{\, /}\;\,p }
\def\ks{\rlap{\, /}\;\,k }

\def\gym{g_{YM}}

\def\adot{\dot{a}}
\def\bdot{\dot{b}}
\def\bpa{\bar{\pa}}

\def\pr{\prime}
\def\ssk{\medskip}
\def\clb{\color{blue}}
\def\clr{\color{red}}
\def\clg{\color{green}}
\def\clp{\color{purple}}
\def\clc{\color{cyan}}
\def\clm{\color{magenta}}
\def\cly{\color{yellow}}

\def\bfA{{\bf A}}
\def\bfB{{\bf B}}
\def\bfK{{\bf K}}
\def\bfU{{\bf U}}
\def\bfX{{\bf X}}
\def\bfY{{\bf Y}}
\def\bfZ{{\bf Z}}
\def\bfg{{\bf g}}
\def\bfn{{\bf n}}

\def\bsk{\bigskip}
\def\ssk{\medskip}

\def\Ec{{\cal E}}

\begin{document}

\overfullrule=0pt
\parskip=2pt
\parindent=12pt
\headheight=0in \headsep=0in \topmargin=0in
\oddsidemargin=0in

\vspace{ -3cm}
\thispagestyle{empty}

 \vspace{0.1cm}

\setcounter{equation}{0}
\setcounter{footnote}{0}
\setcounter{section}{0}

\begin{center}

{\Large\bf Influence of finite-temperature effects on CMB power spectrum}

\vskip 0.8cm

\vspace{0.5cm}


I. Y. Park$^\dagger$ and Peter Y. Wui$\,^{\spadesuit}$
\\

\vspace{0.3cm}

%
{\it {}{$^\dagger$}Department of Applied Mathematics,
Philander Smith University 
                               \\
Little Rock, AR 72202, USA \\
inyongpark05@gmail.com \\

\vspace{.2in}
{}{$\spadesuit$}Department of Business Administration,
University of Arkansas at Pine Bluff
                               \\
Pine Bluff, AR 71601, USA\\
wuiy@uapb.edu
}

 \vspace{.5cm}

\end{center}

 \vspace{0.1cm}

\begin{abstract}

We explore the implications of finite-temperature quantum field theory effects on cosmological parameters within the framework of the $\L$CDM model and its modification. By incorporating temperature-dependent corrections to the cosmological constant, we extend the standard cosmological model to include additional density parameters, $\Omega_{\L_2}$ and $\Omega_{\L_3}$, which arise from finite-T quantum gravitational effects. Using the Cosmic Linear Anisotropy Solving System (CLASS), we analyze the impact of these corrections on the cosmic microwave background power spectrum and compare the results with the Planck 2018 data. Through brute-force parameter scans and advanced machine learning techniques, including quartic regression, we demonstrate that the inclusion of $\Omega_{\L_2}$ and $\Omega_{\L_3}$ improves the model's fit to the {\em fine} structure of the reference power spectrum. Because $\Omega_{\L_2}$ and $\Omega_{\L_3}$ originate from finite-temperature quantum (loop) corrections rather than being introduced as free dark-energy components, they are expected to be small a priori; their smallness is thus a prediction of the framework, not a sign that they are negligible. The improvement of the fit is measured by the coefficient of determination $R^2$ (the fraction of variance in the parameter--distance relation captured by the regression), together with lower mean squared error and lower AIC/BIC scores than those of the $\L$CDM model. Despite identified methodological limitations, these findings establish an exploratory framework for incorporating finite-temperature quantum corrections into precision cosmology and open new avenues for data-driven exploration of cosmological parameters.

\end{abstract}
\newpage





\section{Introduction}

Understanding the cosmological parameters that govern the Universe is central to modern cosmology \cite{Weinberg,Dodelson}, with models of cosmological parameter estimation evolving to incorporate increasingly sophisticated methods and data sources. Among the most influential studies, the Planck 2018 results \cite{Planck:2018vyg} have provided a comprehensive analysis of cosmological parameters derived from CMB anisotropies, utilizing high-precision data to set a benchmark for subsequent analyses. These measurements impose stringent constraints on key parameters such as the density of different components of the Universe, the Hubble constant, and the spectral index. As a result, they have guided our understanding of the Universe's composition and evolution, and provided a solid foundation for both theoretical models and observational research. However, while these results have enhanced our ability to probe the cosmological landscape, the models themselves are still evolving. For instance, there are ongoing efforts to refine the assumptions underlying the cosmological constant (CC) and the potential roles that quantum effects may play over cosmic time. While the conventional hydrodynamic approach incorporates temperature via classical thermodynamics, it is crucial to also consider the temperature effects from finite-temperature (finite-T) quantum field theory (QFT) (see \cite{Laine:2016hma} for a review), as recently elaborated in references \cite{Park:2021ohu} \cite{Park:2021vro}, and \cite{Park:2024kfn}. Given the high temperatures in the early Universe, these effects are unlikely to be negligible when studying its history. This work aims to analyze the implications of a finite-T-corrected cosmological parameters in light of these considerations.

The conventional hydrodynamic approach as well as the one based on Boltzmann's equations is essentially a classical framework: it does not take into account the quantum effects introduced by finite-T QFT.\footnote{One might argue that quantum effects are already incorporated through the use of field-theoretic free energy. However, this formulation does not provide an adequate framework for the problem at hand; see section 2.1 for a detailed discussion.} In addition to classical thermodynamics, finite-T QFT effects should play a significant role in the cosmological constant, feeding into it through perturbative quantum corrections. Ideally, a second-quantized description whose effective description is provided by the hydrodynamics should be employed. The conceptual basis for such a combination - in particular, why
the off-shell vacuum-bubble contributions and the on-shell
hydrodynamic sector can be added without double-counting - is
provided in section~2.1 by analogy with Weinberg's external-field
analysis of QED (Ch.~13.6 of \cite{Weinberg1}), where the
topological separation between ladder-type diagrams and
vacuum-bubble diagrams plays the central role. Imagine obtaining the quantum action, i.e., 1PI action. (Only the first several terms in the derivative expansion will matter.) The loop contributions to the CC depend on the temperature. This in turn implies that in the corresponding FRLW-type cosmology the CC comes to be time-dependent through the time-dependence of the temperature \cite{Park:2024kfn}.\footnote{A time-dependent CC implies the following for the conservation laws \cite{Park:2024kfn}. (See \cite{Fritzsch:2012qc} and the references therein for earlier related works.) In general, one may not impose separate conservation condition for each component. The conservation equation of the total stress tensor follows from the Einstein equation. This bears an interesting implication for renormalization of the coupling constants.} 
Finite-T effects were considered earlier \cite{Park:2021ohu}\cite{Park:2021vro}\footnote{These works are finite-temperature extensions of earlier works on foliation-based gravity quantization \cite{Park:2014tia}, with one key element being the physical state condition \cite{Park:2015qxa}; see Witten's work \cite{Witten:2018lgb} for a similar idea. In \cite{Park:2015xoa}, the gauge $K=0$ (more generally $K=K_0$), where $K$ denotes the trace of the second fundamental form, was employed as a crucial ingredient for quantization. A similar approach was applied in Witten's recent work \cite{Witten:2022xxp}.} to tackle the long-standing cosmological constant problem. See \cite{Sola:2013gha},   \cite{SolaPeracaula:2022hpd}, and \cite{Moreno-Pulido:2023ryo} for an inspiring review of the CC problem and subsequent works. The CC problem was originally observed from the technical feature of onshell renormalization that the one-loop correction is enormously larger than the observed value of the CC, which necessitated a fine-tuned cancellation between the renormalized value of the CC and the one-loop correction. However, what has been noted in \cite{Park:2021ohu,Park:2021vro,Park:2024kfn} (see also \cite{Ryskin:2014pva} and \cite{Balazs:2022anb}) was that if one takes the renormalized mass to be on the order of the temperature - which is hinted by the fact that the presence of temperature yields a $T^4$-order CC term at the quantum level, the one-loop correction becomes small and this naturally allows one to take a small value of the renormalized value of the CC. It has been proposed that the CC problem may well be a matter of how to manage perturbation theory and finite renormalization, instead of a genuine problem. See also \cite{Sola:2013gha,SolaPeracaula:2022hpd} and \cite{Ageeva:2024qie} for ealier and more recent related discussions, respectively.

We consider a temperature-dependent CC (and its implications for the other cosmological parameters) for several reasons. First, the temperature effects are unavoidable due to perturbative quantum gravitational influences, making it impossible to omit unless negligible. Since we're considering periods around or before recombination, these finite-T QFT effects are not negligible, though small, especially given the observed smallness of the present-day CC. Second, these effects exhibit the characteristics needed for early dark energy (EDE) (see \cite{Poulin:2018cxd} for a review of early dark energy), making the temperature-dependent CC a form of early dark energy present throughout the Universe's history. Their contribution is significant in the radiation-dominant era and around last scattering.

At the phenomenological level, the crux of our approach is that the finite-T effects naturally introduce additional parameters with which to fit the data. It is not difficult to see how these extra parameter arises. As discussed in the body, the form of the CC after taking the leading finite-T quantum correction into account is $\L_{tot}=\L_1+\L_2 \fr{1}{a^4}+\L_3 \fr{1}{a^2}+\cdots$ where $a$ denotes the scale parameter; $\L_1,\L_2$, and $\L_3$ are genuine (viz., time-independent) constants.\footnote{Throughout, the phrase ``temperature-dependent (equivalently, time-dependent) correction to the cosmological constant'' is used as convenient shorthand. Strictly, $\L_1$, $\L_2$, and $\L_3$ are genuine constants, and what acquires time dependence is the total vacuum-energy term $\L_{tot}(t)$ through the scale-factor dependence of the finite-temperature contributions. 
We use this shorthand in the same loose but helpful way the term such as "Schwarzschild vacuum" or "monopole vacuum" (which refers to a space that is not actually empty) is used.} A natural question is whether the coefficients $\Lambda_1$, $\Lambda_2$, and $\Lambda_3$ 
are fixed by the underlying model parameters and therefore cannot be treated as free. 
This is not the case: in quantum field theory, these coefficients are subject to 
renormalization conditions and are not determined a priori by the Lagrangian parameters 
alone. The situation is directly analogous to mass renormalization in elementary QFT 
textbooks: the renormalized mass is kept as a free parameter throughout the calculation 
and is fixed only at the final stage by imposing a renormalization condition. 
In the same way, $\Lambda_1$, $\Lambda_2$, and $\Lambda_3$ remain free until physical 
renormalization conditions are specified, and their values are precisely what we 
constrain observationally in the present work. In view of this it is natural to introduce the corresponding density parameters $\Omega_{\L_2},\Omega_{\L_3}$ in addition to $\Omega_{\L_1}$, the usual vacuum energy density parameter. (One may raise that the parameters $\Omega_{\L_2}$ and $\Omega_{\L_3}$ can be absorbed into the $\Omega_{r}$ and $\Omega_{K}$, the radiation and curvature density parameters, respectively; see the discussion below for this.)
At the early stage of the matter dominance, the $\L_{tot}$ term, which is now temperature-dependent and scales as $\sim \fr1{a^4}$, is not negligible compared with the $a^{-3}$ behavior of matter. In other words, the $\Lambda_{tot}$ term is a priori expected to make a non-negligible contribution to the Hubble constant between radiation dominance and radiation-matter equality, the period important for some early dark energy models. We thus consider extension of the original $\L$CDM model by including $\Omega_{\Lambda_2}$ and $\Omega_{\Lambda_3}$. For the actual analysis of the finite-T modified system, we employ the cosmic linear anisotropy solving system (CLASS) \cite{Blas:2011rf, CLASS home}. 

More specifically we consider the model that has the following eight parameters,\\
$(\Omega_{\Lambda_2},\; \Omega_{\Lambda_3},\; h,\;\w_b,\;\w_{cdm},\;A_s,\;n_s,\;\t_{reio})$ \cite{mod_CLASS}, and compare it with the standard $\L$CDM model with $(\Omega_{K},\;h,\;\w_b,\;\w_{cdm},\;A_s,\;n_s,\;\t_{reio})$ in the following manner. For suitable ranges of the parameters, the corresponding power spectra can be generated via CLASS, which can then be plotted to yield the corresponding curve. Each curve being the discrete power spectrum over the multipole moment parameter $\ell$, its Euclidean distance to the Planck experimental curve can therefore be defined. By direct scan it is possible to determine the values of the cosmological parameter set that yields the minimum distance to the Planck 2018 curve. 

While the Planck analysis focused on observational data with standard Bayesian methods, the present work employs statistical and machine learning (ML) regression techniques, specifically quartic regression, to compare the predictive accuracy and generalizability of the two models. Recent work has highlighted the growing role of machine learning and information-theoretic methods in extracting physical insight from complex cosmological and astrophysical models. For example, Piras et al. \cite{Piras:2022wso} developed a robust mutual-information-based framework to interpret latent representations in deep learning models and connect them directly to physically meaningful parameters. Their study demonstrates that modern statistical and machine learning tools can be used not merely for prediction but also for interpretable inference in high-dimensional physical systems. Motivated by this broader shift toward interpretable data-driven modeling in cosmology and astrophysics, the present work adopts a complementary machine-learning-assisted approach to cosmological parameter estimation. Rather than relying exclusively on Bayesian sampling of parameter space, we construct surrogate regression models that map cosmological parameters to observable quantities and use these models to explore the phenomenological implications of finite-temperature quantum corrections to the cosmological constant.

In the standard Bayesian parameter inference framework adopted by the 2018 Planck collaboration, the observed $C_\ell$'s are treated as realizations of random variables, while the cosmological parameters are modeled as random variables endowed with prior distributions. This is a well-established Bayesian statistical paradigm. In contrast, the current ML-enhanced deterministic approach treats the cosmological parameters as input variables in a polynomial regression model. (See the introduction of section 3 for the reason and justification for adopting a different approach.) The coefficients of the polynomial - which relate these parameters to the angular power spectrum - are determined through regression. Subsequently, the values of the cosmological parameters that minimize the discrepancy with the 2018 Planck power spectrum curve are identified. Although we are employing frequentist framework, this is not a standard frequentist approach: the cosmological parameters are promoted to input variables in the regression setup, while the coefficients of the polynomial - rather than the cosmological parameters themselves - assume the role of parameters in the statistical sense. The resulting model may be viewed as a surrogate model that approximates the mapping from cosmological parameters to observables. 

Both $(\Omega_{\Lambda_3},\; h,\;\w_b,\;\w_{cdm},\;A_s,\;n_s,\;\t_{reio})$- and $(\Omega_{\Lambda_2},\; \Omega_{\Lambda_3},\; h,\;\w_b,\;\w_{cdm},\;A_s,\;n_s,\;\t_{reio})$- models demonstrate exceptional accuracy, with the values of the coefficient of determination $R^2$ nearing 99.9 \% for training- as well as testing- dataset. This metric, with mean squared error (MSE), highlights the model's ability to accurately capture the underlying variance of the distance variable while maintaining low error rates. The results are then compared to the CLASS default model and the Planck 2018 model.

\vspace{.2in}

The rest of the paper is organized as follows. In section 2, we begin by recalling the key results from the work \cite{Park:2024kfn} on the finite-temperature quantum gravity (QG) effects. Section 2.1 outlines the scope of the present work using a simple setup. We explain why the finite-temperature vacuum energy parameters 
$\Omega_{\Lambda_2}$ and $\Omega_{\Lambda_3}$ are not redundant with the radiation 
and curvature densities, despite having the same scale-factor dependence inside the Hubble parameter $H$. Their origin lies in quantum loop corrections to 
the cosmological constant, and they enter the effective action differently from 
real radiation or geometric curvature. The non-degeneracy of these parameters has been explicitly verified using a modified version of CLASS. Also, we distinguish the present framework from quantum field-theoretic thermodynamics formulated through Helmholtz free energy. Specifically, the former is offshell, while the latter is inherently onshell. In section 2.2, we adopt Weinberg's approach \cite{Weinberg}, essentially reproducing the power spectrum plot derived therein, and compare it with the Planck 2018 results \cite{Planck:2018vyg}. Although Weinberg's formalism is not suitable for a detailed comparison with current data, it serves as a useful starting point. This approach aids in understanding the underlying physics at intermediate steps and also provides an analytic expression for the final CMB power spectrum, $\frac{\ell(\ell+1)}{2 \pi} C_\ell$. Following this, we examine the temperature-corrected cosmological constant (CC). Using Mathematica, we compute the values of the cosmological parameters, modified by the presence of the time-dependent CC, that best fit the fiducial Planck 2018 results. Many of the steps from Chapters 6 and 7 of \cite{Weinberg} can be carried over with minimal modifications. For clarity, we demonstrate the analysis by including $\Omega_{\L_2}$ but omitting $\Omega_{\L_3}$ to keep the Mathematica calculations manageable in terms of memory usage. In section 3, we begin by highlighting how our approach differs significantly from that of Planck 2018, and we explain the rationale behind this distinction. We first establish baseline results using the curved $\Lambda$CDM model with spatial curvature $\Omega_K$. This 7-parameter framework provides a reference for evaluating the performance of quantum gravitational extensions. For computation of the extended models, we employ a quantum-modified version of the CLASS \cite{mod_CLASS}, which includes the additional parameters $\Omega_{\Lambda_2}$ and $\Omega_{\Lambda_3}$, and obtain the best fit for the 2018 Planck data \cite{Planck:2018vyg}. We consider two models: one with $\Omega_{\Lambda_3}$ alone and one with both $\Omega_{\Lambda_2}$ and $\Omega_{\Lambda_3}$, extending our earlier Mathematica analysis in which $\Omega_{\Lambda_3}$ was absent.
Initially, we perform a brute-force parameter scan, revealing that the theoretical power spectrum is sensitive to both $\Omega_{\Lambda_2}$ and $\Omega_{\Lambda_3}$, along with the other six parameters. Given the inherent randomness in the subsequent sampling methods adopted, we validate these results using an independent method. Specifically, we apply various machine learning techniques, including random forest and others, to assess the predictive performance of both the 7-parameter and 8-parameter models in estimating the distance variable. We use quartic regression - which performed best among the methods we tested - to optimize the feature values and minimize the predicted distance. Model selection across polynomial degrees is guided by the Akaike Information Criterion (AIC) and the Bayesian Information Criterion (BIC), both of which penalize model complexity to guard against overfitting. The quantum gravitational models achieve dramatically lower AIC and BIC values than the standard $\L$CDM baseline - with AIC improving from $+471$ to approximately $-3138$ for the 7-parameter model - indicating that the improvement in fit cannot be attributed to the additional parameters alone.  Consistency with the phenomenological early-dark-energy bounds obtained in \cite{Gomez-Valent:2021cbe} is checked. To evaluate the robustness and reliability of the regression-based parameter estimates, we employ bootstrap resampling, a nonparametric method that complements conventional inference techniques. The inclusion of $\Omega_{\Lambda_2}$ in the 8-parameter model proves to be valuable, as it improves the accuracy of the predictions while maintaining strong generalizability. We end the section by discussing (un)naturalness of the value $\Omega_{\L_2}$ obtained.
Finally, in section 4, we conclude with a summary and potential directions for future research. A summary of the statistical and machine learning terminology used in this work is provided in Appendix A.

\section{Finite-T effects on cosmological parameters}

Before turning to the numerical analysis in section~3, we clarify several conceptual and technical aspects underlying the finite-temperature modifications considered here. We review the origin and physical interpretation of the additional parameters
$\Omega_{\Lambda_2}$ and $\Omega_{\Lambda_3}$, explain why they represent genuinely distinct contributions rather than redundancies of existing cosmological components,
and motivate the structure of the modified cosmological equations that follow.

The discussion is organized as follows. Section~2.1 examines a simpler systems to build intuition for the origin and non-redundancy of $\Omega_{\Lambda_2}$ and $\Omega_{\Lambda_3}$. We begin with a scalar field theory at finite temperature in a flat background, which illustrates how quantum loop
corrections generate temperature-dependent contributions to the cosmological constant at one-loop. We then consider an Einstein-Hilbert action with a cosmological constant
to demonstrate explicitly the non-redundancy of $\Omega_{\Lambda_3}$, followed by an
analogous argument for $\Omega_{\Lambda_2}$. These examples are deliberately simple; the finite-temperature analyses in sections~2.2 and~3 work with the more realistic
settings of hydrodynamic and kinetic matter systems, respectively, coupled to gravity
with a cosmological constant shifted by the temperature-dependent contributions.

The physical basis for these parameters comes from a recent finite-temperature
one-loop renormalization analysis of the Standard Model coupled to gravity~\cite{Park:2024kfn}, which shows that the coupling constants - most notably the cosmological constant - acquire temperature-dependent, and therefore
time-dependent, quantum corrections. The two leading contributions scale as $T^4$
and $T^2$, giving rise to $\Omega_{\Lambda_2}$ and $\Omega_{\Lambda_3}$, respectively. Since the temperature scales as $T \sim a^{-1}$, the associated energy-density components scale in the same way as radiation and spatial curvature;
nevertheless, as shown in section~2.1, they are not redundant - their origins, physical interpretations, and roles in the effective action are distinct from those of $\Omega_r$ and $\Omega_K$. Section~2.1 also explains, by analogy with Weinberg's
external-field analysis of QED in Ch.~13.6 of \cite{Weinberg1},
why combining the off-shell vacuum-bubble contributions with the
on-shell hydrodynamic/kinetic sector does not constitute
double-counting. Section~2.2 opens with a brief review of Weinberg's
semi-analytic formalism \cite{Weinberg} to set the stage for the subsequent numerical analysis, before turning to $\Omega_{\Lambda_2}$ and examining its impact on the standard
cosmological parameters within that framework.

\subsection{On non-redundancy and relevance of $\Omega_{\L_2}$ and $\Omega_{\L_3}$}

Finite-temperature effects feed the CC through perturbative loops. The CC thus becomes temperature-dependent for an FLRW background. One might mistakenly assume that finite-temperature quantum loop-induced contributions to the cosmological constant, such as those involving photons and neutrinos, are already included as part of the radiation contribution. However, this view overlooks the role of virtual particles in the cosmological constant. The contributions from Standard Model (SM) particles (and gravitons) arise from the vacuum energy of virtual particles, while radiation corresponds to the energy density of physical, onshell particles moving through the Universe.

Let us first briefly review how a matter or metric field generically produces quantum corrections to the cosmological constant. For simplicity, we consider a scalar field, but a parallel analysis can be carried out for any matter field or for the graviton. We consider a flat background with curved space generalization outlined afterwards. Normally, a field-independent term - such as the cosmological constant - is disregarded in a flat background. However, once gravity is included, such a term takes a diffeomorphism-covariant form and appears multiplied by a factor of $\sqrt{-g}$, where $g_{\mu\nu}$ denotes the metric. The same is true for finite-temperature corrections in a flat background; such term must be kept track of.

Consider the Einstein-scalar system:
\bea
S_{\text{ren}} = -\int d^4x \left[ \frac{1}{2} \partial_\mu \phi \partial^\mu \phi + \frac{1}{2} m^2 \phi^2 + \frac{\lambda}{4!} \phi^4 \right].
\eea
We define
\bea
V_{\text{ren}} \equiv \frac{1}{2} m^2 \phi^2 + \frac{\lambda}{4!} \phi^4, \quad m_\phi^2 \equiv m^2 + \frac{1}{2} \lambda \phi^2.
\eea
At zero temperature, one-loop effective action is discussed in many textbooks; for instance see chap. 16 of \cite{Weinberg2}. The result contains a constant term, the CC. The finite temperature analogue is also well known: it can be written as (see, e.g., \cite{Arnold:1992rz} and the review in \cite{Laine:2016hma})
\bea
V_{\text{one-loop}}(\phi,T) = V_{\text{ren}} + \hbar J(m_\phi,T), \quad \text{where} \quad
J(m_\phi,T) = \;\;\mathclap{\displaystyle\int}\mathclap{\textstyle\sum}\;\;\; \frac{1}{2} \ln(P^2 + m_\phi^2).  \la{Voneloop}
\eea
where $P=(\omega_n,{\bf p})$. This leads to the expression
\bea
J(m_\phi,T) = -\frac{\pi^2}{90} T^4 \!+\! \frac{m_\phi^2}{24} T^2 \!-\! \frac{1}{12\pi} m_\phi^3 T \!-\! \frac{m_\phi^4}{2(4\pi)^2} \left[ \ln\left( \frac{\bar{\mu} e^{\gamma}}{4\pi T} \right) \!+\! \frac{1}{2\epsilon} \right] \!+\! \frac{\zeta(3)}{3(4\pi)^4} \frac{m_\phi^6}{T^2} + \dots   \la{JmT} 
\eea
where \(\ln \mu^2 = \ln(4\pi \mu^2) - \gamma\) is the renormalization scale, \(\gamma\) is the Euler constant, \(\epsilon\) is the regularization parameter, and  \(\zeta(3)\) denotes the value of the Riemann zeta function at argument $3$. Note that \(\hbar\) is explicitly displayed. The factor $\hbar$ multiplies $J(m_\phi,T)$ as a whole, marking the entire $J$ as the one-loop contribution, with $V_{\text{ren}}$ in \rf{Voneloop} the classical (tree-level) potential. All of \rf{JmT} -- not only the manifestly $T$-dependent terms -- is thus part of this single one-loop correction: the $T$-independent $m_\phi^4$ log term and $1/\epsilon$ pole are the zero-temperature one-loop (Coleman--Weinberg) vacuum contribution, so $J(m_\phi,T)$ should not be read as an entirely thermal object. What has been shown in \cite{Park:2024kfn} is that the corresponding calculation in an FLRW background can be done with a diffeomorphism covariant result. In particular, one has the same form of $J(m_\phi,T)$ now with time-dependent temperature, $T\sim \fr1{a}$. The density components associated with $\Omega_{\L_2}$ and $\Omega_{\L_3}$ arise from the first two terms in \rf{JmT}. If one considers the SM, then each field has its own $J$; enumeration of the numerical coefficients of $T^4$ was carried out therein; also see section 3 below. As for the zero-temperature limit, since \rf{JmT} is a high-$T$ expansion in $m_\phi/T$, the limit is taken not term by term but at the level of the renormalized $J$: as shown in \cite{Park:2024kfn}, the UV counterterms retain their covariant, $T$-independent zero-temperature forms even at finite $T$, so the $T$-independent part of $J$ is precisely the renormalized zero-temperature one-loop (Coleman--Weinberg) effective potential.

With this preliminary, let us discuss the non-redundancy of $\Omega_{\L_2}$ and $\Omega_{\L_3}$. Fluctuations in the finite-temperature contributions, $\delta\rho_{\Lambda_2}$ and $\delta\rho_{\Lambda_3}$, are neglected, as they are both quantum in origin and parametrically suppressed, making the variations subleading. Consequently, their impact on perturbations is minimal, and the parameters $\Omega_{\Lambda_2}$ and $\Omega_{\Lambda_3}$ enter effectively only through the modified Hubble parameter $H$. This statement holds at the level of approximation adopted in the present analysis; at higher orders, additional contributions such as $\Lambda_{\mathrm{tot}} R_{\mu\nu} R^{\mu\nu}$ are expected to arise. By contrast, the curvature and radiation components, $\Omega_K$ and $\Omega_r$, contribute directly to the fluctuation equations and play an active role in the evolution of cosmological perturbations. In this work, $\Omega_{\Lambda_2}$ and $\Omega_{\Lambda_3}$ are treated phenomenologically, with their inclusion supported by consistency with the numerical results.

The non-degeneracy can be easily seen in the case of $\Omega_{\L_3}$. To illustrate the basic idea in the simplest possible setting, consider the Einstein–Hilbert action with a cosmological constant: 
\bea
{\cal L}&=& \fr1{\k}(-2\L_1+R).
\eea
One expands the metric 
\bea
g_{\m\n}\equiv h_{\m\n}+g_{0\m\u}(K)
\eea
around the background solution denoted by $g_{0\m\n}(K)$, where $K$ denotes the background curvature parameter,
\bea
ds^{2} = a^{2}\left[ d{\bf x}^{2} + K \frac{({\bf x} \cdot d{\bf x})^{2}}{1 - K {\bf x}^{2}} \right].  \la{Kmet}
\eea
Substituting $g_{\m\n}=g_{0\m\n}+h_{\m\n}$ into the action and taking $h_{\m\n}$-variation one gets the field equation. The parameter $K$ then appears in multiple places in the resulting equations,\footnote{For the present discussion (and the analysis in the introduction of section 3) we keep the curvature parameter $K$ nonzero; for the analyses of the extended models in section 3.1 and 3.2, $K$ is set to zero to keep the numerical analysis manageable.} which is why the density parameter $\Omega_K$ enters the standard cosmological equations in several ways, given        
\bea
\Omega_K\equiv -\fr{K}{a_0^2H_0^2}.
\eea
The parameter, $\Omega_{\L_3}$, however, has an entirely different origin. When finite-temperature one-loop corrections are taken into account, the cosmological constant $\Lambda_1$ is shifted schematically as
\bea
\L_{tot}\equiv \L_1+\L_2 \fr{1}{a^4}+\L_3 \fr{1}{a^2}+\cdots.
\eea
The shift arises from the metric analogues of \rf{JmT}. The linearization steps above will have to be repeated with $\L_{tot}$ replacing $\L_1$; the density parameter $\Omega_K$ will appear in multiple places in the standard cosmological equations for the same reason as before. Unlike $\Omega_K$, however, the parameter  $\Omega_{\L_3}$ appears only in the cosmological-constant term of the offshell one-loop effective action. (As mentioned before, this is true at the level of approximation adopted in this work.
At higher orders, terms such as \(\Lambda_{\text{tot}}\, R_{\mu\nu} R^{\mu\nu}\) will arise.) After the standard linearized analysis of a system of gravity coupled with hydrodynamics matter, it enters the Hubble parameter as in \rf{HmodL2q} below, wherein $\Omega_{K}$ is set to zero. Otherwise, $\Omega_{\Lambda_3}$ and $\Omega_{K}$ would appear only in the combination $\Omega_{\Lambda_3} + \Omega_{K}$ in $H$. Nevertheless, these parameters are clearly not redundant: their origins, physical interpretations, and appearances in the action are distinct.  The non-degeneracy of these parameters has also been explicitly verified by running a modified version of CLASS. See Appendix A of \cite{Hatefi:2025lqe}.

Although non-redundancy of the newly introduced parameters should be obvious from the discussion above, let us also consider $\Omega_{\L_2}$ for completeness. To be specific, let us consider a system described by Boltzmann kinetic theory, such as a photon gas, coupled to a scalar field - e.g., the Higgs field - on a curved spacetime background. The total action is given by
\bea
S = S_{EH} + S_{sf} + S_{kin}, \la{sfkin}
\eea
where $S_{EH}$ denotes the Einstein-Hilbert action with a CC, $S_{sf}$ is the action for the scalar field, and $S_{kin}$ accounts for the kinetic system. The explicit form of $S_{kin}$ is known in specific cases (see, e.g., \cite{Schutz:1970my,Carter:1992im}), such as for an ideal gas, but will not be required for the present discussion. The corresponding metric field equation contains the well-known expression for the stress-energy tensor:
\bea
T^{\mu\nu} = \frac{1}{\sqrt{-g}} \sum_n \delta^3(x_n(t) - x)\, p_n^\mu(t)\, p_n^\nu(t)\, E_n(t).
\eea
To convey the idea without unessential complications, we consider a fixed, non-dynamical background metric. Also, we focus on the $T^4$ scaling associated with $\Omega_{\L_2}$. The standard treatment focuses solely on the classical thermodynamics of the kinetic system, which yields the familiar exact $T^4$ dependence for the energy density, as noted, for example, in footnote 1 of section 3.1 in \cite{Weinberg}.

In contrast, the present work amounts to examining quantum field-theoretic effects arising from the scalar field sector $S_{sf}$ (and other SM particles and graviton, which do not change the qualitative behavior) at finite temperature. Evidently, such a scalar field cannot be viewed as radiation due to their heavy mass, and they contribute to the CC as virtual particles, i.e., through vacuum bubble diagrams. Those contributions take a form of an infinite series in powers of the temperature, the leading term of which scales as $T^4$ as reviewed above.

As seen in \eqref{Voneloop}, \(J(m_\phi,T)\) represents a quantum correction, in contrast to the classical kinetic theory derivation of thermodynamic relations, such as \(p = \frac{1}{3} \rho\), mentioned in \cite{Weinberg}. The kinetic theory treatment is purely classical and does not involve quantum corrections. Therefore the offshell field-theoretic vacuum energy contributions are distinct from contributions from the onshell kinetic matter components. A lingering question may remain, given that thermodynamic relations, such as \(p = \frac{1}{3} \rho\), can also be derived in a quantum field-theoretic thermodynamic framework by introducing the Helmholtz free energy. In such a framework, \(V_{\text{one-loop}}(\phi,T)\) is identified as the Helmholtz free energy density, and thermodynamic relations such as \(p = \frac{1}{3} \rho\) can be derived for a massless case by setting the free energy density \(f = \frac{\pi^2}{90} T^4\), so conducting the field theoretical analysis, such as, vacuum loop computation, in the present of the kinetic matter system as in \rf{sfkin} might appear redundant. Such a framework, while useful in certain contexts, is not entirely suitable for the present discussion, as we explain now.

There are several important issues with applying this derivation directly to the cosmological context:

\begin{enumerate}
    
    \item On-shell Condition:
    For the derivation of \(p = \frac{1}{3} \rho\), the scalar field \(\phi\) must be set to zero, i.e., \(\phi = 0\), which is a (trivial) solution of the field equation. This implies an {\em onshell} analysis, where the field \(\phi\) is fixed to a solution. However, this is not the objective of the present work. The current analysis keeps the \(\phi\)-dependent part in order to obtain the quantum-corrected offshell aciton, namely the 1PI action, and focuses on the \(\phi\)-independent contributions to the cosmological constant. Thus, the field-theoretic derivation of \(p = \frac{1}{3} \rho\) in the quantum field-theoretic thermodynamics does not imply that, for example, the \(T^4\) contribution from vacuum diagrams is redundant to the radiation contribution.
    
    \item Massive Particles in the Standard Model:
    In the Standard Model (SM), there are massive particles that cannot be regarded as radiation. These particles contribute to the \(T^4\)- and \(T^2\)-terms in the cosmological constant, even though they do not fit within the classical radiation picture (i.e., \(p = \frac{1}{3} \rho\)).
\end{enumerate}

The correct context for conducting field-theoretic on-shell
thermodynamics, and the conceptual justification for combining the
field-theoretic vacuum-bubble contributions with the
hydrodynamic/kinetic-theory matter sector without double-counting,
is best stated by analogy with Weinberg's external-field
approximation in QED (Ch.~13.6 of \cite{Weinberg1}). There, a $U(1)$
gauge system is analyzed in the presence of a heavy charged
particle. The relevant diagrams are the ladder-type graphs
(Fig.~13.5 of \cite{Weinberg1}) in which photon lines connect the
light electron line to the heavy-particle line. Weinberg shows that
the sum of these ladder diagrams factorizes and is equivalent to a
set of Feynman rules in which the light electron interacts with a
classical external vector potential - equivalently, to adding an
interaction term $\mathcal{L}_{\rm ext}(x) = \mathcal{A}_\mu(x)
J_e^\mu(x)$ (Eq.~(13.6.14) of \cite{Weinberg1}) to the Lagrangian.
The heavy particle's quantum-level
interaction with the light electron through the ladder diagrams is
repackaged as an effective external-field interaction. Crucially,
vacuum-bubble diagrams are topologically distinct from these ladder
graphs -- they do not connect to the heavy-particle line -- and
therefore contribute separately to the effective action.

The analogous procedure for the SM coupled to gravity proceeds as
follows. After spontaneous symmetry breaking, one isolates, say, the
$U(1)$ sector (the photon) and carries out the finite-temperature
analogue of Weinberg's analysis, now in the presence of radiation. The
analysis analogous to that of the ladder-diagram should produce the usual
hydrodynamic part of the effective action, with the macroscopic
radiation density $\rho_r$ emerging as the counterpart of
Eq.~(13.6.14) - i.e., the on-shell content that feeds $\Omega_r$ in
the Friedmann equation. Vacuum-bubble diagrams - including those of
the $U(1)$ sector itself, as well as those of the heavy SM fields
(Higgs, top quark, electroweak gauge bosons) - are topologically
disjoint from the ladder diagrams and contribute separately to the
cosmological-constant sector of the effective action. These
vacuum-bubble contributions are what
$\Omega_{\Lambda_2}$ and $\Omega_{\Lambda_3}$ parametrize at leading
order. Because ladder-type and vacuum-bubble diagrams form disjoint
topological classes, adding them does not double-count the $U(1)$
sector, even when one includes the vacuum bubbles of the photon
itself. 

Furthermore, for matter components like dust (e.g.\ stars, galaxies), a purely
classical treatment is appropriate: such objects are not described
by Feynman diagrams but by kinetic theory, so the only sensible
description is via kinetic theory or hydrodynamics. Treating all
on-shell matter components uniformly via the kinetic-theory /
hydrodynamic sector is therefore the appropriate setup, while the
off-shell vacuum-bubble content of the SM (and the graviton) is
collected into the cosmological-constant sector parametrized by
$\Omega_{\Lambda_1}$, $\Omega_{\Lambda_2}$, and
$\Omega_{\Lambda_3}$.

To even further clarify the distinction between $(\Omega_{r},
\Omega_K)$ and $(\Omega_{\L_2}, \Omega_{\L_3})$, let us consider the
most general system:
\bea
S = S_{EH} + S_{SM} + S_{kin}, \la{sfkin}
\eea
and its quantization. The Weinberg-style derivation just outlined is
the rigorous first-principles route. In practice, however, it is
more convenient to consider a slightly different but conceptually
simpler approach which, within the typical approximations used in
cosmology, arrives at the same starting point. This approach begins
by formulating an effective field theory for gravity and SM
particles, without including the kinetic-theory sector. (The term
``effective'' refers here to an effective field theory derived by
keeping only the leading terms in the derivative expansion of the
1PI action.) This effective action accounts for quantum corrections,
including a $T^4$ term (and a $T^2$ term) in the CC sector.
Subsequently, a kinetic or hydrodynamic system - which includes
matter components such as $\r_r$ or $\r_{cdm}$ - can be coupled to
this effective field theory. For a purely hydrodynamic system, all
SM fields can be set to zero.

\vspace{.3in}

One might ask whether such finite-temperature shifts in the cosmological constant can be phenomenologically relevant, given their quantum origin. There are several reasons to expect that they can. First, the observed value of the cosmological constant is already extremely small, so there is no \emph{a priori} reason to assume that quantum corrections must be negligible relative to it. Second, the physical importance of quantum corrections in gravitational systems is well established, for example in black hole physics, where they lead to qualitatively new structures such as quantum-originated accretion disks and jets \cite{Nurmagambetov:2020ann}. Most importantly, since the temperature scales as 
\bea
T \sim \fr{1}{a}
\eea                                                                   
the leading finite-temperature correction scales as $T^4$, making it potentially significant in the early Universe. Moreover, the values of the effective cosmological parameters are influenced by cumulative finite-temperature effects over cosmic history. The present analysis demonstrates explicitly that these effects induce nontrivial and phenomenologically relevant shifts in the cosmological parameters.

\subsection{Finite-T modifications implemented in Weinberg's approach}

With the role of the leading T-dependent part of the vacuum energy in dealing with the CC problem understood, one can consider the finite-T contributions for the present purpose.\footnote{Here and in the numerical analysis in section 3 we are implicit about the renormalization scheme. We postpone the discussion until it becomes relevant, i.e., section \ref{OL2na} where (un)naturalness of the value of $\Omega_{\L_2}$ obtained is discussed.} Let us now consider conducting cosmological perturbation analysis with the 1PI action (with only the first few leading-order terms kept) coupled with the hydrodynamic (or Boltzmann) system. One of the implications of \cite{Park:2021ohu,Park:2021vro,Park:2024kfn} is that the leading quantum corrections yield the following form of the time-dependent vacuum energy\footnote{Since higher-order loop corrections are among the terms represented by $(\cdots)$, additional time-dependent terms with different $a$-scalings will arise. These terms are not considered in the present work, although they will be addressed in the conclusion.}
\be
\L_{tot}(t)=\L_1+ \L_2\left(\fr{a_0}{a}\right)^4 
+ \L_3\left(\fr{a_0}{a}\right)^2 +\cdots    \la{tdcc}
\ee
where $\Lambda_1, \Lambda_2$ are time-independent constants, with $\Lambda_1 + \Lambda_2 + \Lambda_3 + \cdots$ representing the known present value of the CC. The time-dependent CC in eq. \rf{tdcc} introduces additional density parameters, $\Omega_{\L_2}$ and $\Omega_{\L_3}$, corresponding to the $a^{-4}$- and $a^{-2}$-scalings, respectively, in addition to the usual density parameter $\Omega_{\L_1}$. Although the CC term associated with $\Omega_{\L_2}$ scales as $1/a^4$, it does not correspond to radiation, as extensively discussed in section 2.1. Similarly, $\Omega_{\L_3}$, which appears in the off-shell action through quantum corrections, is conceptually distinct from $\Omega_K$, which is associated with the parameter $K$ in the on-shell solution.

\subsubsection{Angular power spectrum without finite-T effects}

In this section, as a warm-up, we study the implications of the presence of \\
$\Omega_{\L_2}$ for $(h,\;\w_b,\;\w_{cdm},\;A_s,\;n_s,\;\t_{reio})$ by adopting and extending Weinberg's semi-analytic formalism. The reason for employing the Weinberg's formalism prior to a more precise numerical analysis by CLASS is that the former has the analytic final expression for the CMB power spectrum, $\fr{\ell(\ell+1)}{2 \pi} C_\ell$. The Weinberg's approach captures physics of the processes involved. By considering both approaches one can grasp intuition as well as accuracy.

We begin with a lightning review CMB angular power spectrum computation in \cite{Weinberg}. There are two main components in the analysis therein: the fractional ionization $X$ and the power spectrum itself $C_{TT,l}^S$ given in \rf{capX} and \rf{cttps} below, respectively. The subscripts $TT$ denote temperature-temperature correlation and $S$ scalar.
The fractional ionization $X$  is defined as 
\be
X=\fr{n_e}{n}, \quad n=n_p+n_H. 
\ee
where $n_e,n_p,n_H$ denote the density of electrons, protons, hydrogens, respectively. It obeys 
\be
	\fr{dX}{dT}=\frac{n \alpha}{T H}
	\left(1+\frac{\tilde{\b}}{\Gamma_{2s}+\frac{8 \pi  H}{\lambda_\a^3 n (1-X)}}\right)^{-1}
	\left(X^2-\frac{1-X}{S}\right)  \la{capX}
	\ee
where
\bea
H&=& 7.204\times 10^{-19}\, T^{\frac{3}{2}} \sqrt{1.523 \times 10^{-5}\, T+h^2 \left(\frac{2.725}{T}\right)^3 \Omega_ {\Lambda_1}+h^2 \,\Omega_M},\nn\\
\alpha &=& \frac{1.4377 \times 10^{-10} \,T^{-0.6166}}{5.085 \times 10^{-3}\, T^{0.53}+1}\; cm^3 s^{-1},\quad
\tilde{\b} = 2.4147 \times 10^{15}cm^{-3} \; e^{-39474/T}\, T^{\frac{3}{2}}\, \alpha , \nn\\
\Gamma_{2s}&=&8.22458 \,s^{-1},\quad \l_\a =1215.682\times 10^{-8} cm,\quad
S=1.747 \times 10^{-22} \; e^{157894/T}\, T^{\frac{3}{2}}  \Omega_B  h^2.  \nn\\
\eea	
The $h^2 \left(\frac{2.725}{T}\right)^3 \Omega_ {\Lambda_1}$ term in $H$ is small and can be disregarded \cite{Weinberg}. We have used $\tilde{\b}$, instead of $\b$ in \cite{Weinberg}, to avoid confusion with the integration variable $\b$ appearing in \rf{cttps} below.
This result is obtained with a commonly adopted assumption that the radiation is viewed to suddenly go from thermal equilibrium with matter to a free expansion at the red shift $z_L$. The quantatative definition of the corresponding temperature $T_L$ is based on the so-called visibility function associated with the fractional ionization. To calculate $T_L$, consider the opacity
\be
{\cal O}=1-\exp\Big[-\int_{t(T)}^{t_0}c\s_{\cal T} n_e dt\Big]=1-\exp\Big[-c\s_T\int_{T_0}^{T}  dT' \fr{n_e(T')}{H(T')T'}\Big]  \la{symbO}
\ee
where $\s_{{\cal T}}=0.66525 \times 10^{-24} \; cm^2$, and the visibility function (it is usually defined in terms of $z$, $g(z)$.) 
\be
g(T)= \fr{d}{dT}\exp\Big[-c\s_T\int_{T_0}^{T}  dT' \fr{n_e(T')}{H(T')T'}\Big]
     = \fr{d}{dT}\exp\Big[-c\s_T\int_{T_0}^{T}  dT' \fr{n(T')X(T')}{H(T')T'}\Big] \la{goT}
\ee
where $n_e=nX$ was used in the second equality. The explicit form of $n$ is 
\be
n=0.76\; \fr{3H_0^2 \Omega_B}{8\pi G m_p}\Big(\fr{T}{T_{\g 0}}\Big)^3
\ee
The last scattering temperature $T_L$ is defined by
\be
\fr{dg}{dT}\Big|_{T=T_L}=0. \la{dgdT}
\ee
As for $C_{TT,\ell}^S$, after extensive analysis it was shown in \cite{Weinberg} that the power spectrum formula is given by (in some equations in this section $\t_{reio}$ is denoted by $\t_{reion}$ by following the conventions of \cite{Weinberg}):
\bea
\fr{l(l+1)}{2\pi} C_{TT,l}^S &=& \fr{4\pi}{25}T_0^2N^2e^{-2\t_{\small{{reion}}}}
\int_1^\infty d\b \left(\fr{\b l}{l_{{\cal R}}}\right)^{n_S-1}  \nn\\
&&\hspace{-.9in} \left\{
\fr{3(\b^2-1)^{\fr12}}{\b^4 (1+R_L)^{\fr32}}e^{-2\b^2 l^2/l_D^2} {\cal S}^2(\b l/ l_T)\sin^2\Big(\b l/l_H+\D(\b l/l_T) \Big)\right. \nn\\
&&\left. \hspace{-1.5in} +\fr1{\b^2(\b^2-1)^{\fr12}}\Big[3{\cal T}(\b l/l_T)R_L- (1+R_L)^{-\fr14} e^{-\b^2 l^2/l_D^2} {\cal S}(\b l/ l_T)\cos\Big(\b l/l_H+\D(\b l/l_T) \Big)\Big]^2
\right\}. \; \la{cttps}
\eea
Above, $N$ is a normalization paramter and $\t_{\small{{reion}}}$ denotes the optical depth of reionized plasma. By following \cite{Weinberg} we set
\be
 \fr{4\pi}{25}T_0^2N^2e^{-2\t_{\small{{reion}}}}=519.7\, \m K^2.
\ee
$R_L$ is defined as the ratio of the background baryon and photon densities
\bea
R_L \equiv\fr34 \fr{{\r_B}}{{\r}_\g}\Big|_{z=z_L} = \fr{3\Omega_B}{4\Omega_\g}  \fr1{(1+z_L)}. \la{RL}
\eea 
The transfer functions ${\cal S}, {\cal T}, \D$ were numerically determined and given by
\bea
{\cal S}(\k) &\equiv& \left(\frac{  5^{1/2} (0.1657 \k)^6 +(0.5116 \k)^4+(1.209 \k)^2+1}{(0.1657 \k)^6+(0.4249 \k)^4+(0.9459 \k)^2+1}\right)^2 \nn\\
{\cal T}(\k) &\equiv& \frac{\log [(0.124 \k)^2+1]}{(0.124 \k)^2} \sqrt{ \frac{(0.2197 \k)^6+(0.4452 \k)^4+(1.257 \k)^2+1}{(0.3927 \k)^6+(0.8568 \k)^4+(1.606 \k)^2+1}}\nn\\
\D(\k) &\equiv&\left(\frac{(0.2578 \k)^6+(0.5986 \k)^4+(1.1547 \k)^2}{(0.2204 \k)^8+(0.4581 \k)^6+(0.8707 \k)^4+(1.723 \k)^2+1}\right). 
\eea
The definitions of $l_T,l_D,l_H,l_{{\cal R}}$ are as follows:
\bea
l_T \equiv \fr{d_A}{d_T},\quad l_D \equiv \fr{d_A}{d_D},\quad  l_H \equiv \fr{d_A}{d_H},\quad l_{\cal R} \equiv (1+z_L)k_{{\cal R}} {d_A}
\eea
where the arbitrary pivot scale $k_{\cal R}$ is chose to be $=0.05$, and $d_T$, the acoustic horizon distance at last scattering $d_H$, and the angular diameter distance $d_A$ are given by
\bea
d_T &\equiv& \k \fr{a_L}{q}=\fr{\sqrt{\Omega_R}}{(1+z_L)H_0 \Omega_M} \nn\\
d_H&\equiv&\fr{2}{H_0 \sqrt{3R_L \Omega_M}\;(1+z_L)^{3/2}}\ln\Big(\fr{\sqrt{1+R_L}+\sqrt{R_{EQ}+R_L}}{1+\sqrt{R_{EQ}}}\Big) \nn\\
d_A &\equiv& r_L a_L= \fr1{H_0}\fr1{1+z_L}\int_{(1+z_L)^{-1}}^1 
\fr{dx}{\sqrt{\Omega_\L x^4+\Omega_M x}}
\eea
respectively. The symbols $\k, R_{EQ}$ are defined as
\bea
\k\equiv \fr{\sqrt{2} \;q}{a_{EQ}H_{EQ}},\quad R_{EQ}=\fr34 \fr{{\r_B}}{{\r}_\g}\Big|_{z=z_{EQ}}.
\eea
 For damping distances one has, 
\be
d_D^2=d_{\mbox{Landau}}^2+d_{\mbox{Silk}}^2
\ee
where, for Landau damping,
\bea
d_{\mbox{Landau}}^2&=&\fr{3\s^2 t_L^2}{8T_L^2(1+R_L)}
\eea
where $\s$ denotes the standard deviation of ${\cal O}'(T)$ when approximated as a Gaussian distribution, 
and 
\bea
d_{\mbox{Silk}}^2=\fr{c R_L^2}{6(1-Y)n_{B0}\,\s_{{\cal T}}\, H_0\, \sqrt{\Omega_B}\,R_0^{9/2}}\int_0^{R_L}\fr{R^2 dR}{X\,(1+R)\sqrt{R_{EQ}+R}}\Big(\fr{16}{15}+\fr{R^2}{1+R}\Big) \la{dsilkeq}
\eea
for Silk damping. The paramter $t_L$ in the Landau damping can be computed by the general formula for the time for an event with the red shift $z$
\bea
t(z)=\fr{1}{H_0}\int_0^{\fr1{1+z}}\fr{dx}{x} \Big[\Omega_{\Lambda}+\Omega_M x^{-3}  +\Omega_{R}x^{-4}\Big]^{-\fr12}.   \la{toz}
\eea
Before delving into the finite-temperature corrections in the next section, we would like to briefly comment on some of the approximations made in the analysis of \cite{Weinberg}. In addition to the approximation of treating recombination as nearly instantaneous, several other approximations and simplifications were made. For example, the integrated Sachs–Wolfe effect, which primarily affects small values of $\ell$, was neglected. Additionally, the tight coupling approximation was used for the early photon-baryon fluid. (The CLASS model, in contrast, accounts for these effects with greater accuracy.) Furthermore, although less significant, there are minor "discrepancies" between the results obtained by Weinberg and those presented in the present work. For instance, consider the last two rows in Table 2.3 of \cite{Weinberg}. When using the same data from that table, the values for the standard deviations turned out to be slightly different in our analysis. This indicates that there is a small program-dependence in the computation of sigma.

\subsubsection{Angular power spectrum with finite-T effects}

For Mathematica analysis we only include $\Omega_2$, but not $\Omega_3$, to keep the analysis manageable. The paramter $\Omega_3$ will be included in section 3 where the analysis by employing CLASS is carried out. The presence of an additional density parameter, $\Omega_{\L_2}$, manifests itself only through the Hubble parameter $H$. Let us enumerate the channels through which the additional density parameter, $\Omega_{\L_2}$, modifies various cosmological parameters.
First, the presence of $\Omega_{\L_2}$ modifies the fractional ionization $X$ through the modified $H$,
\bea
H= 7.204\times 10^{-19}\, T^{\frac{3}{2}} \sqrt{1.523 \times 10^{-5}\, T+ \left(\frac{2.725}{T}\right)^3 \Omega_ {\Lambda_1}h^2+\Omega_M h^2  +\frac{T}{2.725} \Omega_{\Lambda_2} h^2}. \la{HmodL2}
\eea
The modification of X leads, via eq. \rf{dgdT}, to a modified value of $z_L$, thus other quantities, such as $R_L$, as well. The power spectrum formula eq. \rf{cttps} was obtained by first obtaining a semi-analytic solution -  the analysis carried out in Ch. 6 of \cite{Weinberg} - and substituting it into a two-point correlator of the temperature variation. The analysis of Ch. 6 of \cite{Weinberg} can be carried over with the minor change of $H$, eq. \rf{HmodL2}, whenever it appears. As for $C_{TT,l}^S$ in eq. \rf{cttps}, one can use ${\cal T, S},\; \Delta$ without modifications. The finite-T corrections modifies $d_A$ and $d_D,d_T,d_H,l_{\cal R}$ as well through the modification of the former:
\bea
d_A =\frac{2.9979\times 10^{3}}{ (1+z_L)\sqrt{ \Omega_{ \Lambda_1} h^2 x^4+\Omega_M h^2 x + (\Omega_{\Lambda_2}+\Omega_R)h^2 }}.
\eea

\ni Fig. 1 shows sample plots of $X$ for different values of $\Omega_{\L_2}$.
\begin{figure}
	\hspace{-.3in}
	\centerline{
		\begin{minipage}[b]{7cm}
			\epsfxsize=8cm
			\epsfbox{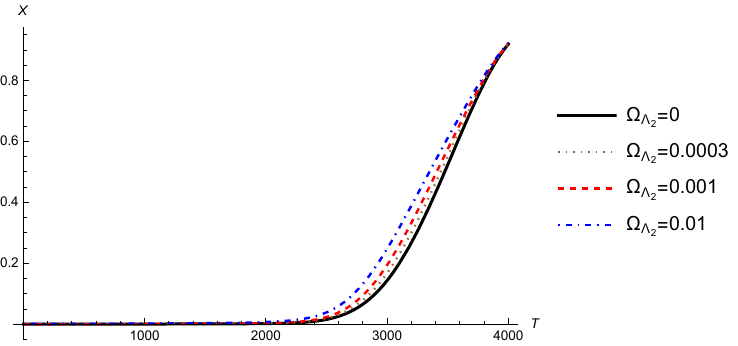}
		\end{minipage}
	}
	\caption{typical behaviors of the fractional ionization $X$ as a function of $\Omega_{\L_2}$; four arbitrary values of $\Omega_{\L_2}$ are chosen for demonstration}
	\label{gfsquare}
\end{figure}
Although $X$ decreases slower toward lower temperature with increased values of $\Omega_{\L_2}$, explicit analysis based on eq. \rf{dgdT} reveals that the finite-T effect actually increase the value of $T_L$.

\begin{figure}
	\hspace{-.1in}
	\centerline{
		\begin{minipage}[b]{7cm}
			\epsfxsize=7cm
			\epsfbox{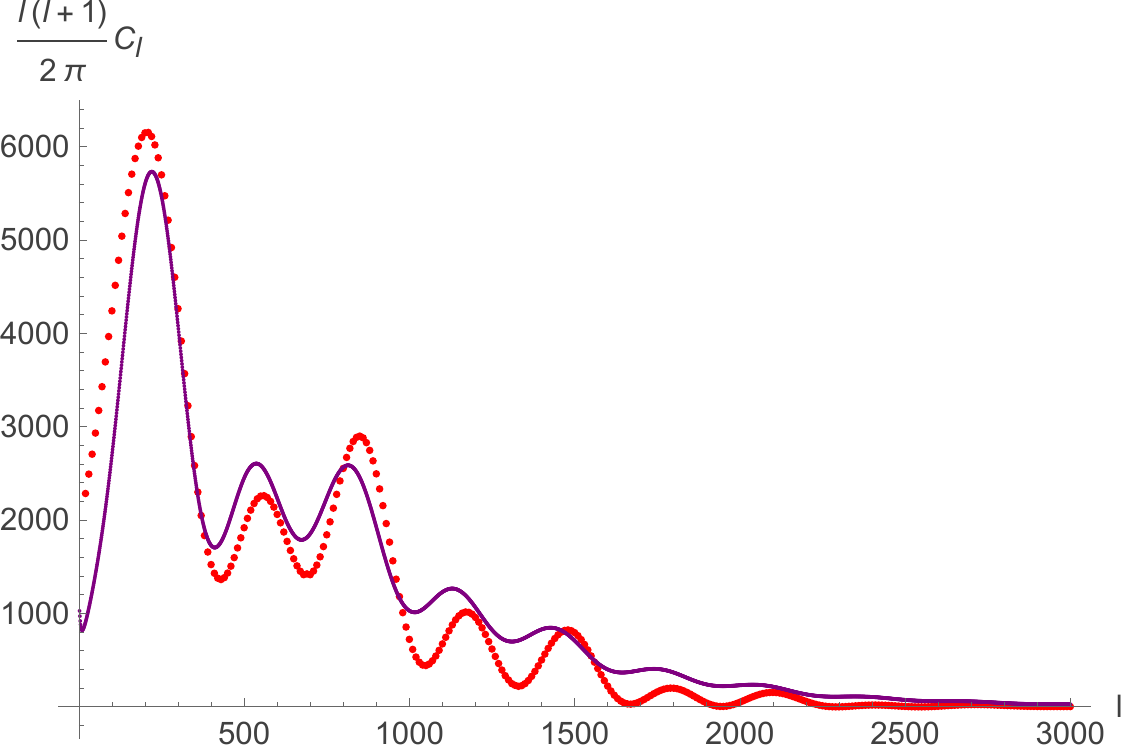}
		\end{minipage}
	}
	\caption{$(h,\Omega_B,\Omega_{\L_2})$-interpolation with $\Omega_M=0.13299/h^2$: the best fit parameters are $ (h,\Omega_B,\Omega_{\L_2},n_s)=(0.785508, 0.0409387, 0.000100, 0.97881)$.}
	\label{gfsquare}
\end{figure}

To obtain the CMB power spectrum in terms of the seven parameters—six usual ones,
\bea
h,\Omega_B,\Omega_M,N,n_S,\t_{\small{{reio}}},
\eea
plus $\Omega_{\L_2}$ - one must first derive the various parameters that appear in the angular power spectrum in eq. \rf{cttps}, such as $l_T, l_D, l_H, l_{{\cal R}}$, as functions of the seven parameters. To keep the Mathematica numerical study manageable, we adopt the values of $N$ and $\tau_{\text{reio}}$ from \cite{Weinberg} and focus on how the parameters depend on four cosmological parameters, $(h, \Omega_B, \Omega_M, \Omega_{\L_2})$. The determination of the optimal $n_s$ can be done at the final stage. In other words, $n_s$ can be addressed after evaluating $C_{TT,l}^S$, once the dependencies of $l_T, l_D, l_H, l_{{\cal R}}$ on the independent cosmological parameters are determined. Our main effort focuses on determining the dependencies of $l_T, l_D, l_H, l_{{\cal R}}$ on the four parameters $(h, \Omega_B, \Omega_M, \Omega_{\L_2})$.

The expressions for $l_T, l_D, l_H, l_{{\cal R}}$ themselves involve parameters such as $z_L, \sigma, R_L, R_{EQ}, t_L$ (where $\sigma$ denotes the standard deviation of ${\cal O}'(T)$, where ${\cal O}$ is given in eq. \rf{symbO}, when approximated as a Gaussian distribution). These parameters need to be expressed in terms of $(h, \Omega_B, \Omega_M, \Omega_{\L_2})$ in order to quantitatively determine the values of the four parameters for the best fit.

Although it would be ideal to express these parameters in terms of the full set $(h, \Omega_B, \Omega_M, \Omega_{\L_2})$, this turns out to be a highly memory-intensive process that requires more powerful computational resources (as was the case for $\sigma$, for instance). We will not pursue this full task in the present work. Instead, we grouped the parameters into two sets: $(h, \Omega_B, \Omega_{\L_2})$ and $(h, \Omega_M, \Omega_{\L_2})$, and separately study the dependence of the various parameters on each of these two groups.
For each set, $(h, \Omega_M, \Omega_{\L_2})$ and $(h, \Omega_B, \Omega_{\L_2})$, we computed the power spectrum. An example of the resulting plots is shown in Fig. \ref{gfsquare}.\footnote{A minor technical note: the solid line in the plot is based on the CLASS default.ini file, not the Planck 2018 data. As we will see in the next section, the two are visually indistinguishable. We used the CLASS data because its $\ell$-range is larger, while the Planck 2018 data extends only up to $\ell = 2508$.}

Weinberg's approach is insightful and valuable for understanding the physics of key events during the intermediate period between the generation of the CMB and its observation today. However, as noted in his book, its primary aim is not the precise calculation of the power spectrum curve. For such purposes, computational tools like CAMB or CLASS are typically employed.

\section{Modified CLASS and machine learning techniques}

In this section, we detail our approach to parameter inference, which diverges from the standard Bayesian methods employed by collaborations like Planck. Our methodology emphasizes model building over rigorous parameter inference and focuses on constructing a phenomenological model with additional parameters $\Omega_{\L_2}$ and $\Omega_{\L_3}$ that can accurately reproduce the observed cosmic microwave background (CMB) angular power spectrum. The quantities $R^2$, MSE, AIC and BIC reported in this section characterize the accuracy of the quartic surrogate in reproducing the Euclidean distance variable across CLASSy-generated parameter sets; they are not likelihood-based goodness-of-fit statistics for the CMB data. Because our target is the unbinned Planck 2018 best-fit theoretical spectrum, used without its covariance matrix, the distance metric does not define a likelihood and the fit improvements reported below should be read as relative model-building performance rather than as statistical detections or confidence levels.

We perform two complementary analyses: a brute-force parameter scan and a subsequent analysis leveraging statistical and machine-learning techniques. A core component of our approach is the use of a finite-temperature modified version of the CLASS. This allows us to perform almost the entire parameter exploration within its Python wrapper, CLASSy, significantly enhancing computational efficiency.

The primary motivation for our non-standard (i.e., non-Bayesian) approach is the behavior of the newly introduced parameters, especially $\Omega_{\L_2}$. Initial theoretical estimates for its magnitude were found to be several orders of magnitude larger than the values that ultimately provided the best fit to the Planck 2018 data. Using these initial estimates resulted in power spectra that deviated significantly from the observed data. Instead of a standard Bayesian inference, which would be computationally prohibitive due to the vast parameter space and the lack of a reliable prior for $\Omega_{\L_2}$, we adopted an iterative, trial-and-error procedure. This process involved gradually reducing the value of $\Omega_{\L_2}$ and performing extensive parameter scans until a good fit was achieved. Although this method lacks statistical rigor (see the conclusion for further discussion), it proved to be an effective strategy for finding the correct order of magnitude for this parameter.

For our analysis, we used the unbinned theoretical comparison spectrum file from the Planck 2018 collaboration:\\ \texttt{COM\_PowerSpect\_CMB-base-plikHM-TTTEEE-lowl-lowE-lensing-minimum-theory\_R3.01.txt}.\\
It is important to note that this "cleaned" dataset, as far as we know, 
does not come with a corresponding covariance matrix, which further justified our departure from standard likelihood-based methods that rely on such information.\footnote{
The file \texttt{COM\_PowerSpect\_CMB-base-plikHM-TTTEEE-lowl-lowE-lensing-minimum-theory\_R3.01.txt}
contains the unbinned best-fit theoretical power spectrum of the $\Lambda$CDM model. As a theoretical prediction, it is a smooth curve with values tabulated at every 
integer multipole~$\ell$. In contrast, the measured Planck spectra are provided in binned form, for example in \texttt{COM\_PowerSpect\_CMB-TT-full\_R3.01.txt}. 
These binned spectra are reported at non-integer effective multipoles, and their correct use requires the corresponding covariance matrix to account for cosmic variance, instrumental noise, and residual foregrounds. The unbinned 
theoretical spectrum, on the other hand, serves as a clean, noise-free reference for direct 
comparison.}

A key advantage of our methodology is its computational efficiency. The entire parameter search process is performed within CLASSy, with only a small, nearly instantaneous external code required to calculate and sort distance values. In contrast, a standard Bayesian analysis would necessitate additional software packages like MontePython or Cobaya. Our tests showed that integrating these packages can increase runtime by a factor of six to eight. Given that our data collection alone took over three months, this efficiency was critical. While a standard Bayesian approach might require a supercomputer for a similar task, our method provides a more accessible and rapid way to explore this complex parameter space. Once our trial-and-error method establishes a rough order of magnitude for $\Omega_{\L_2}$, a more focused scan using tools like Cobaya becomes a feasible and less time-consuming next step.

In the modified version of CLASS \cite{mod_CLASS}, the original Hubble constant $H(t)$ is replaced by the following that contains the finite-T corrections\footnote{The details of modification can be found in the documentation in \cite{mod_CLASS}.}:
\bea
H= 7.204\times 10^{-19}\, T^{\frac{3}{2}} \sqrt{1.523 \times 10^{-5}\, T+ \left(\frac{2.725}{T}\right)^3 \Omega_ {\Lambda_1}h^2+\Omega_M h^2  +\frac{T}{2.725} \Omega_{\Lambda_2} h^2   +\frac{2.725}{T} \Omega_{\Lambda_3} h^2 } \la{HmodL2q}. \nn\\
\eea
Once the values of the cosmological parameters are fixed, the power spectra are computed using the modified CLASS. A large number of parameter sets can be scanned by running the CLASS Python wrapper, CLASSy. Each resulting spectrum is then fed into a code that calculates the Euclidean distance between the generated curve and the Planck data \cite{Planck:2018vyg}, \texttt{COM\_PowerSpect\_CMB-base-plikHM-TTTEEE-lowl-lowE-lensing-minimum-theory\_R3.01.txt}. By these comprehensive scans and distance calculations we obtain tens of millions of data points.
The details of the brute-force scan will vary slightly each time because the data collection procedure involves random sampling. It will be nice to check the qualitative features of the results by an independent method. For this purpose we employ machine learning techniques and evaluate the predictive performance of the models in estimating the distance.

Given that in sectoin 2.2 we analyzed the effects of $\Omega_{\Lambda_2}$ in the absence of $\Omega_{\Lambda_3}$, in this section we consider two models: one including $\Omega_{\Lambda_3}$ only, and the other including both $\Omega_{\Lambda_2}$ and $\Omega_{\Lambda_3}$: the first is the modified $\Lambda$CDM model, which has 7 parameters: 
\bea
 (\Omega_{\Lambda_3}, h, \omega_b, \omega_{cdm}, A_s, n_s, \tau_{\text{reio}}). 
\eea
The second model has 8 parameters: 
\bea
(\Omega_{\Lambda_2}, \Omega_{\Lambda_3}, h, \omega_b, \omega_{cdm}, A_s, n_s, \tau_{\text{reio}}). 
\eea
For each model, we use 5000 datasets, selected from a much larger data sample as explained below. We then apply a regression-based machine learning technique to estimate the minimum distance between the generated spectra and the Planck 2018 data.

The objective is to determine which model best explains the variance in the target variable, the distance $d$, while balancing model complexity and performance. Model selection is based on maximizing $R^2$ and minimizing the MSE in both training and testing datasets. Using quartic regression - which proved to be the best-performing among the models we tested,\footnote{Regression likely performs well because the distance can, in principle, be expressed as an analytic function of the cosmological parameters. The regression model effectively approximates its multivariable Taylor expansion.} including random forest and others - we compare the accuracy and generalization capability of both approaches. The comparison shows that including the additional parameters from finite-temperature QG effects modifies the values of the cosmological parameters.

\vspace{.3in}

Before presenting the detailed analysis of quantum gravitational extensions to the $\Lambda$CDM cosmology, we first establish baseline results using the standard $\Lambda$CDM model with spatial curvature, $\Omega_K$. This 7-parameter framework serves as a reference point against which the performance of quantum gravitational parameterizations can be evaluated. The present analysis of the standard $\Lambda$CDM employs the same computational and statistical methodology described in subsequent sections, utilizing 5,000 minimum-distance parameter sets selected from a comprehensive brute-force scan. The dataset exhibits the following statistical characteristics (see Table 1).
\begin{table}[H]
\centering
\small
\begin{tabular}{lcccc}
\toprule
\textbf{Parameter} & \textbf{Mean} & \textbf{Std Dev} & \textbf{Range} \\
\midrule
distance & 64.62 & 9.64 & [30.78, 77.53] \\
$h$ & 0.6792 & 0.0012 & [0.677, 0.682] \\
$\omega_b$ & 0.022378 & 0.000028 & [0.02232, 0.02242] \\
$\omega_{\rm cdm}$ & 0.11984 & 0.00031 & [0.119, 0.12056] \\
$A_s$ ($\times 10^{-9}$) & 2.098 & 0.0046 & [2.090, 2.110] \\
$n_s$ & 0.96595 & 0.00058 & [0.965, 0.967] \\
$\tau_{\rm reio}$ & 0.05391 & 0.00102 & [0.052, 0.056] \\
$\Omega_k$ & $-0.000022$ & 0.000080 & [$-0.0002$, $0.0001$] \\
\bottomrule
\end{tabular}
\caption{Statistical summary of the standard $\Lambda$CDM model parameters}
\label{tab:lcdm_stats}
\end{table}
\ni Notably, the standard $\Lambda$CDM model exhibits a higher minimum distance (30.78) and larger mean distance (64.62 $\pm$ 9.64) compared to what will be observed in quantum gravitational extensions.\footnote{As for the comparison with the unmodified CLASS' default six parameter model, the Euclidean distance between the Planck 2018 reference spectrum and spectra generated by the \texttt{default.ini} settings is found to be
\bea
d_{\mathrm{default}} \simeq 75 .
\eea
(Empirically, spectra with $d \lesssim 1000$ are visually indistinguishable from the Planck reference curve.) With $\Omega_{\Lambda_2}$ and $\Omega_{\Lambda_3}$, the minimal distances obtained are significantly smaller,
\bea
d_{\mathrm{modified}} \simeq 25\text{--}30 
\eea
so the reduction in distance represents a nontrivial quantitative improvement.} The spatial curvature parameter $\Omega_K$ is consistent with zero within uncertainties, with a mean of $-2.2 \times 10^{-5}$, though the dataset explores both positive and negative curvature values.

Polynomial regression analysis identifies degree-4 as optimal across all evaluation criteria ($R^2$, MSE, AIC, BIC). See Appendix A for the statistical glossary. The performance metrics reveal:

\begin{align}
R^2_{\mathrm{train}} &= 0.9924, \nn\\
R^2_{\mathrm{test}}  &= 0.9912, \nn\\
\mathrm{MSE}_{\mathrm{train}} &= 0.7055, \nn\\
\mathrm{MSE}_{\mathrm{test}} &= 0.8281, \nn\\
\mathrm{AIC} &= 471.40, \nn\\
\mathrm{BIC} &= 2090.96.
\end{align}
\ni These values, while representing excellent predictive accuracy, are notably inferior to the performance achieved by the models with $\Omega_{\L_2}$ and $\Omega_{\L_3}$, suggesting that the introduction of quantum corrections may capture cosmological dynamics more effectively than allowing non-zero spatial curvature within the standard framework.

Ten-fold cross-validation confirms model robustness, with mean metrics across folds as follows:
\begin{equation}
\begin{aligned}
\text{Train MSE} &= 0.699 \pm 0.007 \\
\text{Test MSE}  &= 0.858 \pm 0.068 \\
\text{Train } R^2 &= 0.9925 \pm 0.0001 \\
\text{Test } R^2  &= 0.9907 \pm 0.0010.
\end{aligned}
\end{equation}
The minimum predicted distance across cross-validation folds averages $34.71 \pm 1.38$, indicating substantial variability in the optimization landscape. Bootstrap resampling with 1,000 iterations yields the following optimized parameter estimates with 95\% confidence intervals:
\begin{table}[h]
\centering
\small
\begin{tabular}{lcc}
\toprule
\textbf{Parameter} & \textbf{Mean} & \textbf{95\% CI} \\
\midrule
$h$ & 0.6796 & [0.6794, 0.6803] \\
$\omega_b$ & 0.022389 & [0.022387, 0.022409] \\
$\omega_{\rm cdm}$ & 0.11981 & [0.11967, 0.11989] \\
$A_s$ ($\times 10^{-9}$) & 2.0926 & [2.0922, 2.0944] \\
$n_s$ & 0.96604 & [0.96589, 0.96656] \\
$\tau_{\rm reio}$ & 0.05259 & [0.05239, 0.05317] \\
$\Omega_k$ & 0.000042 & [$-0.000011$, 0.000068] \\
predicted distance & 32.78 & [32.24, 33.84] \\
\bottomrule
\end{tabular}
\caption{Bootstrap-derived parameter estimates for standard $\Lambda$CDM}
\label{tab:lcdm_bootstrap}
\end{table}
The narrow confidence intervals for most parameters reflect strong consistency across resampled datasets, though $\Omega_K$ exhibits wider relative uncertainty, crossing zero within its confidence bounds. Parameter optimization targeting minimum distance yields 
\bea
\begin{aligned}
h &= 0.679026 \\
\omega_b &= 0.022385 \\
\omega_{\rm cdm} &= 0.119865 \\
A_s &= 2.0857 \times 10^{-9} \\
n_s &= 0.965907 \\
\tau_{\rm reio} &= 0.050890 \\
\Omega_K &= -4.24 \times 10^{-5}
\end{aligned}
\eea
with a predicted minimum distance of 29.56. This represents the best achievable fit within the standard $\Lambda$CDM framework using this dataset. These baseline results establish the performance benchmark against which the finite-T models will be compared, allowing direct assessment of whether quantum corrections provide statistically significant improvements over the standard cosmological paradigm. As we further discuss below, the superior fit quality (MSE reduction by factor of ~8$\sim$10) despite comparable model complexity (both 7-parameter frameworks have identical numbers of features after polynomial expansion) indicates quantum dynamical effects may be physically more relevant than static spatial curvature for explaining cosmological observations.

\subsection{Brute-force scan of finite-T models by CLASSy}

In the analysis by employing modified CLASS, we first conduct brute-force scans to systematically explore the parameter space for the finite-T cosmological models. By generating discrete power spectra over the multipole moment parameter $\ell$ for suitable ranges of the parameters, their distances from the Planck 2018 experimental curve can be calculated. This approach requires extensive scanning of the parameter space, ultimately determining the set of cosmological parameters that minimizes the distance to the Planck 2018 curve. 

As stated in the beginning, the adoption of a distance-based parameter fit is partly motivated by the challenge of determining an appropriate range for $\Omega_{\Lambda_2}$. Given specific numerical values for $\Omega_{\Lambda_2}$ (along with other parameters), one can efficiently and deterministically compute the distance to the Planck power spectrum curve. This contrasts with the Planck analysis framework - such as that implemented via MontePython - which typically relies on stochastic sampling methods (e.g., MCMC), requiring significantly more computational effort for parameter space exploration. Furthermore, to enhance the robustness and efficiency of our analysis, we incorporate machine learning techniques to guide and refine the parameter search, enabling a more targeted and accelerated convergence toward the best-fit regions.

Using modified CLASS, specifically its Python wrapper CLASSy, we conduct extensive parameter scans, ultimately collecting several tens of millions of datasets. This immense volume of data is necessary to perform various checks and validate the robustness of the results. Based on these preliminary explorations, appropriate parameter ranges are carefully refined. From these final intervals, we collect 400,000 datasets for the 7-parameter case and 300,000 datasets for the 8-parameter case. The number of the datasets for the 8-parameter model was deliberately chosen less than that of the 7-parameter model. This was to bring out the fact that the former performs better in spite of this disadvantage. See the related comments below.

The impact of inclusion of $\Omega_{\L_2}$ is reflected on
the actual minimum values of the distance and the corresponding parameters set.  For the 7-parameter model, the minimum distance of the 400,000 datasets is\footnote{To provide a sense of the accuracy associated with a distance value around 30, consider the following example. Using the CLASS configuration file \texttt{base\_2018\_plikHM\_TTTEEE\_lowl\_lowE\_lensing.ini}, and changing the maximum multipole $\ell$ from 2500 to 2508, yields a computed distance of 56.1 (rounded) when compared to the reference data in \texttt{COM\_PowerSpect\_CMB-base-plikHM-TTTEEE-lowl-lowE-lensing-minimum-theory\_R3.01.txt}. Since \texttt{base\_2018\_plikHM\_TTTEEE\_lowl\_lowE\_lensing.ini} is designed to reproduce the Planck 2018 baseline results, the CLASS-generated spectrum should closely match the Planck theoretical spectrum. The observed discrepancy may arise from factors such as numerical precision, CLASS version, or calibration differences.} 
\be
d=28.122388 
\ee
which occurs at\\

\scalebox{0.8}{\hspace{.5in}
\begin{tabular}{ccccccc} 
   $\Omega_{\L_2}$ & $h$ & $\w_b$ & $\w_{cdm}$ & $A_s$ & $n_s$ & $\t_{reio}$ \\ 
  0.002339   & 0.677333   & 0.022387   & 0.119889   &  $2.098889\times 10^{-9}$   & 0.9660    & 0.054000   \\ 
\end{tabular}\\
}\\

\ni whereas for the 8-parameter model, the minimum of the 300,000 datasets is
\be
d=26.832229
\ee
with\\
\scalebox{0.8}{
\;\;\;\;
\begin{tabular}{cccccccc}
\footnotesize
$\Omega_{\Lambda_2}$ & $\Omega_{\Lambda_3}$ & $h$ & $\omega_b$ & $\omega_{\text{cdm}}$ & $A_s$ & $n_s$ & $\tau_{\text{reio}}$ \\
$-2.073444 \times 10^{-8}$ & 0.004297 & 0.675500 & 0.022394 & 0.119889 & $2.101111 \times 10^{-9}$ & 0.966000 & 0.054444 \\
\end{tabular}
}
\\

\begin{figure}
	\hspace{-.3in}
	\centerline{
		\begin{minipage}[b]{7cm}
			\epsfxsize=7cm
			\epsfbox{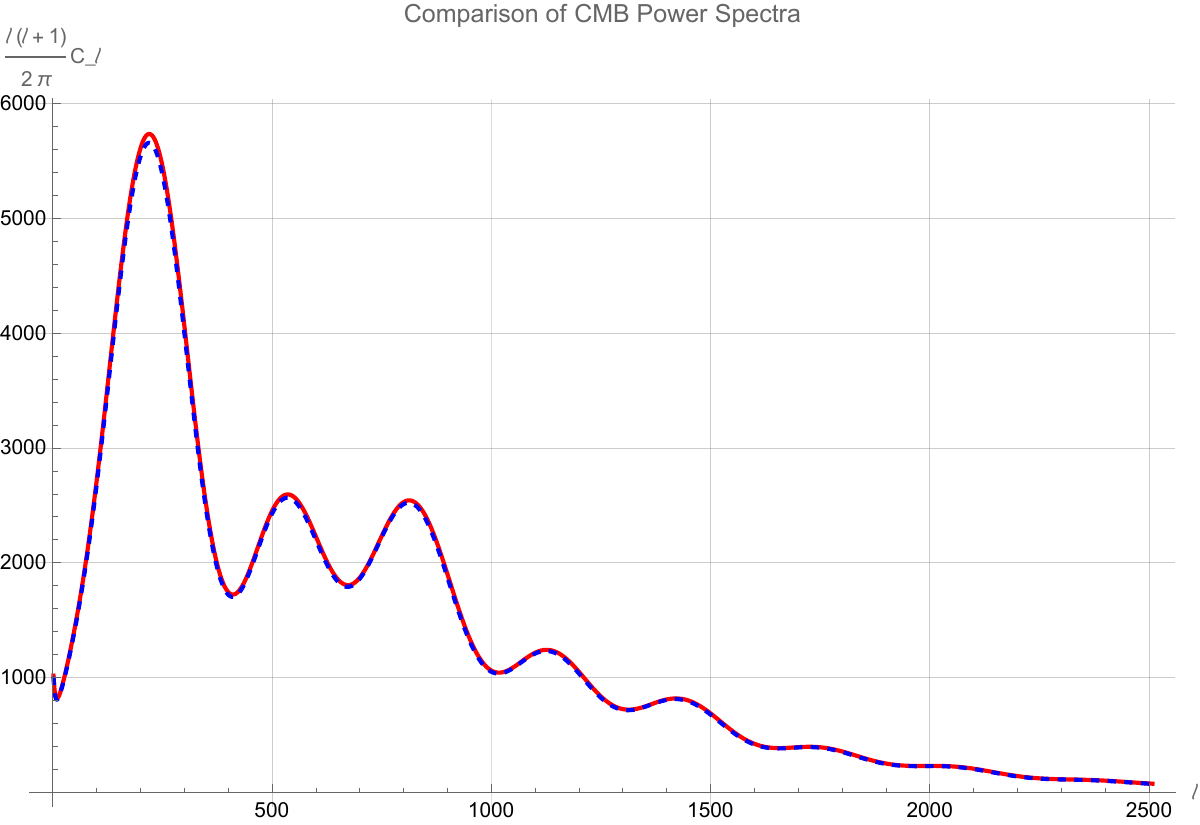}
		\end{minipage}\;\;\;
		\begin{minipage}[b]{7cm}
			\epsfxsize=7cm
			\epsfbox{"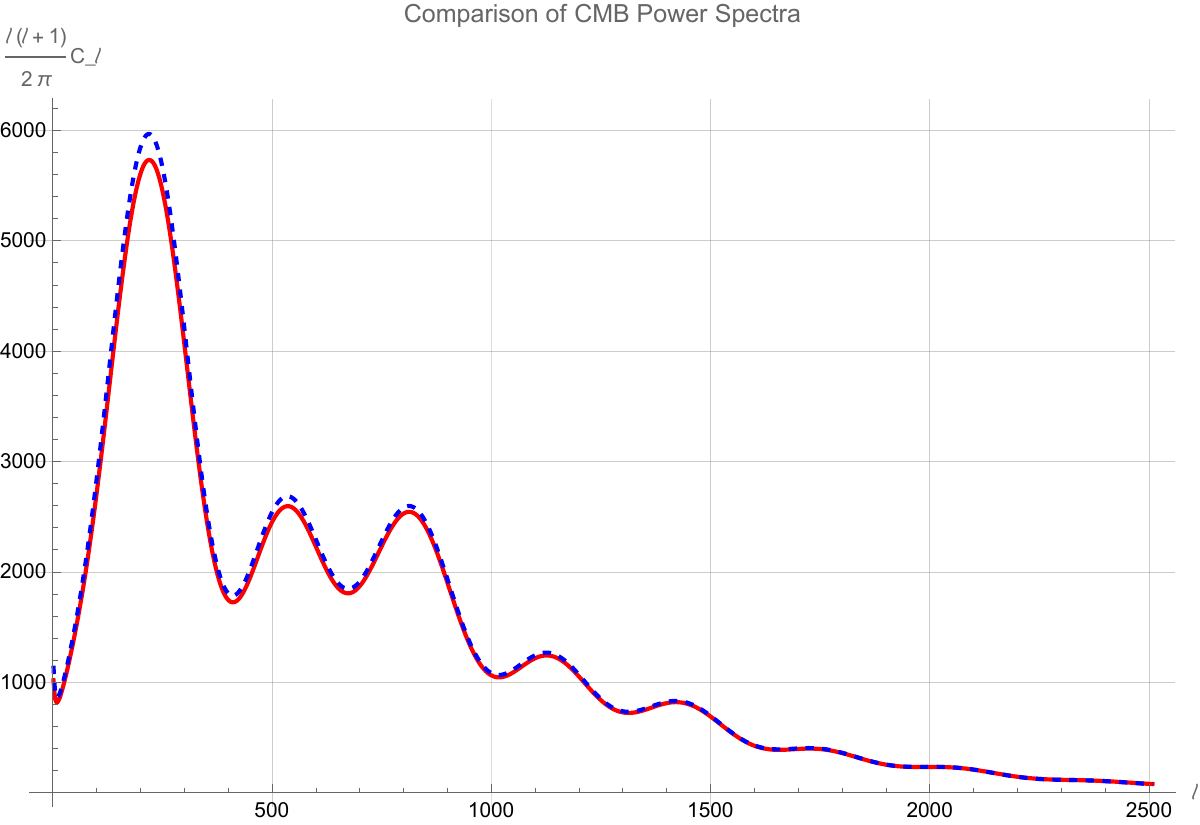"}
		\end{minipage}
	}
\vspace{-.1in}	
\[	\hspace{-.5in}\mbox{(a)} \hspace{2.7in}\mbox{(b)}\]
\vspace{-.3in}
	\caption{The dotted plot in panel (a) represents a curve with a distance $d \approx 1200$ from the Planck 2018 data, while the dotted plot in panel (b) corresponds to a distance $d \approx 3400$.}
	\label{cwsh}
\end{figure}

\ni These minima are actual data values. Note the negative sign of $\Omega_{\L_2}$ to which we will come back later. (See, e.g., footnote \ref{msign}.) Below we obtain the regression functions of these two datasets and obtain the minima predicted by the functions. Those minima will in turn be checked against the values (given in \rf{mdc}) yielded by directly running modified CLASS with the corresponding values of the parameters as input. To provide some context for the distance values, we show two plots in Fig. \ref{cwsh}. As seen in these plots, the curves with distances in the range of several hundreds are nearly identical to the Planck 2018 curve, appearing directly on top of it.

Given that the 8-parameter model includes one additional parameter, a fair comparison would require a dataset larger than 400,000 by some appropriate factor to adequately cover the expanded parameter space. Despite this discrepancy, the results indicate that the minimum distance to the Planck 2018 curve is smaller for the 8-parameter model, an important feature shared by the statistical analysis below.

\subsection{Statistical and machine learning analysis of finite-T models}

Involving random sampling, the specifics of the brute-force scan may vary slightly with each iteration. It would be reassuring to access the qualitative aspects of the results using an independent method. For this, we apply statistical and machine learning techniques to evaluate the predictive performance of the two cosmological models. The minimum distance predicted by each model is explicitly checked by running modified CLASS with the corresponding set of the cosmological parameters; see eq. \rf{mdc} below. Here are some details of the analysis.

The dataset comprises 5,000 observations, i.e., sets of the parameters, for both the 7-parameter and 8-parameter models. To ensure sufficient data for model validation using 10-fold cross-validation, with each fold containing at least 500 samples, the 5,000 minimum-distance data samples are selected from a total of 400,000 and 300,000 CLASSy-generated datasets for the 7-parameter and 8-parameter models, respectively. Each dataset is then split into a training set (70\%), consisting of 3,500 datasets, and a testing set (30\%), consisting of 1,500 datasets. To address differences in the scales of independent parameters, all features are standardized, which transforms each variable to have a mean of 0 and a standard deviation of 1. This mitigates bias toward features with larger ranges and ensures uniform contribution of all variables to the regression analysis.

\begin{table}[h]

\centering

\scriptsize 

\begin{tabularx}{\textwidth}{l|X| *{2}{>{\centering\arraybackslash}X} >{\centering\arraybackslash}X | *{3}{>{\centering\arraybackslash}X}}

\toprule

\multirow{2}{*}{\textbf{Variable}} & \multirow{2}{*}{\textbf{Count}} & \multicolumn{3}{c}{\textbf{7-Variable Model}  (\(\Omega_{\L_3}\), 6 variables)} & \multicolumn{3}{c}{\textbf{8-Variable Model} (\(\Omega_{\L_2}\), \(\Omega_{\Lambda_3}\), 6 variables)} \\

\cmidrule(lr){3-5} \cmidrule(lr){6-8}

& & Mean & STD & Range & Mean & STD & Range \\

\midrule

distance & 5000 & 55.75 & 8.19 & [28.12, 66.92] & 58.80 & 9.70 & [26.83, 71.76] \\

\midrule

\(\Omega_{\Lambda_2}\) & 5000 & - & - & - & -2.07E-08 & 2.32E-11 & \makecell{[-2.075E-08,\\2.068E-08]} \\

\midrule

\(\Omega_{\Lambda_3}\) & 5000 & 0.00233 & 0.000003 & [0.00233, 0.002342] & 0.004289 & 0.00002 & [0.00425, 0.00431] \\

\midrule

\(h\) & 5000 & 0.676876 & 0.000624 & [0.676, 0.678] & 0.675491 & 0.000325 & [0.675, 0.676] \\

\midrule

\(\omega_b\) & 5000 & 0.022376 & 0.000027 & [0.02232, 0.02242] & 0.022393 & 0.000028 & [0.02235, 0.02245] \\

\midrule

\(\omega_{cdm}\) & 5000 & 0.119924 & 0.000207 & [0.11944, 0.12033] & 0.119853 & 0.00018 & [0.11944, 0.12033] \\

\midrule

\(A_s\) & 5000 & 2.10E-09 & 3.58E-12 & [2.09E-09, 2.1056E-09] & 2.10E-09 & 2.98E-12 & [2.0967E-09, 2.1100E-09] \\

\midrule

\(n_s\) & 5000 & 0.965761 & 0.000491 & [0.965, 0.967] & 0.965896 & 0.000487 & [0.965, 0.967] \\

\midrule

\(\tau_{reio}\) & 5000 & 0.053948 & 0.000757 & [0.052, 0.055] & 0.054819 & 0.000591 & [0.054, 0.056] \\

\bottomrule

\end{tabularx}

\caption{Statistical summary of variables in the 7-parameter and 8-parameter models}

\label{table1}

\end{table}

The performance of the polynomial regression models is evaluated using $R^2$, which quantifies the proportion of variance in the dependent variable modeled by the independent variables. Additionally, the MSE used to measure the average squared difference between predicted and actual values, providing a complementary metric to evaluate model performance. Both models achieve exceptional accuracy, with the $R^2$ values nearing 99.9\%. The inclusion of $\Omega_{\Lambda_2}$ in the 8-parameter model introduces additional complexity, which increases the data requirements for achieving stabilization on the learning curve. However, this added complexity ultimately enhances the model's predictive accuracy. Table \ref{table1} presents a detailed statistical summary of the key variables in the 7-parameter and 8-parameter models. Notably, the addition of $\Omega_{\Lambda_2}$ in the 8-parameter model introduces subtle shifts in other variables, such as $h, n_s,$ and $\t_{reio}$. These shifts are accompanied by slightly increased variability in the distance value, as reflected in its higher standard deviation and wider range in the 8-parameter model compared to the 7-parameter model.

We evaluate polynomial regression models of degrees 1 through 7 for both datasets in order to identify an optimal balance between model complexity and predictive performance. The models are built and assessed separately for the 7-parameter and 8-parameter datasets. The evaluation criteria include $R^2$, MSE, the Akaike Information Criterion (AIC), and the Bayesian Information Criterion (BIC). The quartic model was selected as optimal for both configurations. This model captures the relevant nonlinear relationships while avoiding both underfitting and the overfitting tendencies observed in higher-degree polynomials. For the 7-variable model, the degree-4 regression yields
\begin{align}
R^2_{\mathrm{train}} &= 0.99895, \nn\\
R^2_{\mathrm{test}}  &= 0.99878, \nn\\
\mathrm{MSE}_{\mathrm{test}} &= 0.0795, \nn\\
\mathrm{AIC} &= -3138.48, \nn\\
\mathrm{BIC} &= -1385.12 .
\end{align}
For the 8-variable model, the corresponding degree-4 performance metrics are
\begin{align}
R^2_{\mathrm{train}} &= 0.99917, \nn\\
R^2_{\mathrm{test}}  &= 0.99880, \nn\\
\mathrm{MSE}_{\mathrm{test}} &= 0.1096, \nn\\
\mathrm{AIC} &= -2325.93, \nn\\
\mathrm{BIC} &= 304.11 .
\end{align}
The quartic regression model was therefore adopted for all subsequent analyses, including cross-validation, optimization, and bootstrap resampling.

To further assess the robustness and reliability of our regression-based parameter estimates, we apply a non-parametric statistical technique - bootstrap resampling - as a complementary approach to standard inference methods. While our methodology is primarily a model-building exercise rather than a rigorous parameter inference framework, bootstrap analysis offers an empirical means to quantify uncertainty in the optimized parameters and predicted outcomes. 
From the resulting distributions, we compute empirical 95\% confidence intervals (CI) using the 2.5th and 97.5th percentiles, providing a data-driven estimate of sampling variability. These CIs, though not Bayesian credible intervals, serve as practical indicators of the stability and precision of our regression outputs. 

\begin{figure}
	\hspace{-1in}
	\centerline{
		\begin{minipage}[b]{14cm}
			\epsfxsize=18cm
			\epsfbox{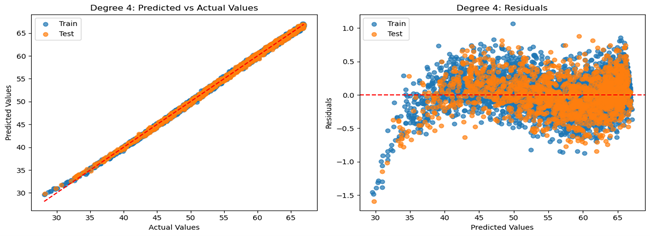}
		\end{minipage}
	}
	\caption{7-parameter model estimation and residuals}
	\label{polyeq7}
\end{figure}

\begin{figure}
	\hspace{-1in}
	\centerline{
		\begin{minipage}[b]{14cm}
			\epsfxsize=18cm
			\epsfbox{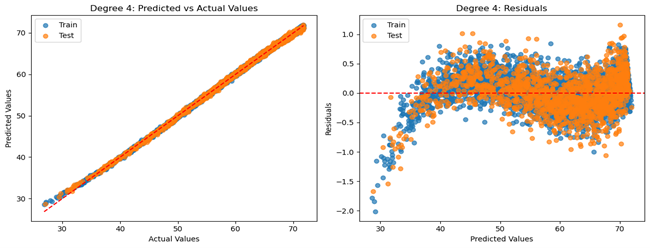}
		\end{minipage}
	}
	\caption{8-parameter model estimation and residuals}
	\label{polyeq8}
\end{figure}

\subsubsection{Single run results}

The $R^2$ metric is calculated separately for the training and testing datasets to evaluate the models' ability to generalize. For the 7-parameter model, the quartic regression achieves an $R^2$ of 99.90\% on the training set and 99.88\% on the testing set, with MSE values of 0.0713 and 0.0795, respectively. In the 8-variable model, including $\Omega_{\Lambda_2}$ improved the training $R^2$ to 99.92\%, though the testing $R^2$ remained at 99.88\%, with MSE values of 0.0792 and 0.1096. While the addition of $\Omega_{\Lambda_2}$ slightly increases the error on the testing set, overall performance remained strong. Figs. \ref{polyeq7} and \ref{polyeq8} visually confirm these results, showing the predicted versus actual values and residual distributions for both models. The plots reveal a near-perfect alignment along the diagonal.

\begin{figure}
	\hspace{-.3in}
	\centerline{
		\begin{minipage}[b]{10cm}
			\epsfxsize=12cm
			\epsfbox{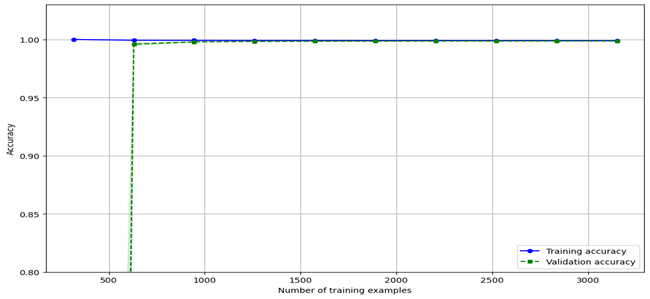}
		\end{minipage}
	}
	\caption{Learning curve of 7 parameter model}
	\label{7learning}
\end{figure}

\begin{figure}
	\hspace{-.3in}
	\centerline{
		\begin{minipage}[b]{10cm}
			\epsfxsize=12cm
			\epsfbox{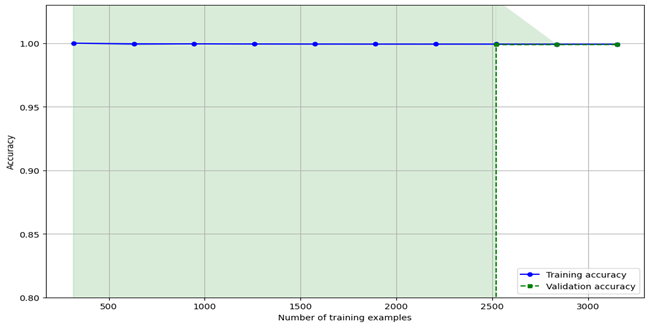}
		\end{minipage}
	}
	\caption{Learning curve of 8-parameter model}
	\label{8learning}
\end{figure}

The learning curves for the 7-parameter and 8-parameter models, shown in Figs. \ref{7learning} and \ref{8learning}, provide valuable insights into how training and validation accuracy evolve as the number of training samples increases. Both models ultimately achieve high accuracy, with training and validation accuracies converging near 99\%, highlighting the effectiveness of the quartic regression approach in predicting the distance variable. For the 7-parameter model, stability is reached  around 1,000 training 
samples, where both accuracies consistently approach 99\%. In contrast, the 8-parameter model, which includes the additional parameter $\Omega_{\Lambda_2}$, requires a significantly larger dataset to stabilize. Stability is achieved at approximately 2,500 training samples, with both accuracies converging near 99\%. This extended stabilization period reflects the added complexity introduced by the extra parameter, which necessitates a larger dataset for effective generalization. Despite the inclusion of $\Omega_{\Lambda_2}$, key parameters such as $A_s$ and $\w_b$ remain highly stable across both models, demonstrating their robustness to changes in model complexity.

\vspace{.1in}
\ni {\bf Minimum distance prediction and check}

\ni The regression results show that the 7-parameter and 8-parameter models predict minimum distance values of 27.88 and 24.39 (both rounded), respectively; see Table \ref{table2}. We again note that the sign of $\Omega_{\Lambda_2}$ is negative; the same sign is observed in \cite{Hatefi:2025lqe}, where its physical significance is discussed. While the 8-parameter model achieves a lower predicted minimum distance, the inclusion of $\Omega_{\Lambda_2}$ in the 8-parameter model led to adjustments in other feature values, most notably $\Omega_{\Lambda_3}$, which influenced the predicted distance outcome. To validate these predictions, we calculated the distances for the corresponding parameters values; the actual CLASS-computed distance values turn out to be
\bea
\begin{tabular}{|c|c|} 
  \hline 
  distance of 7-parameter model & distance of 8-parameter model  \\ 
  \hline 
  26.03   & 24.95     \\ 
  \hline 
\end{tabular}
\la{mdc}
\eea
where the values have been rounded. This result  confirms that the 8-parameter model achieves a slightly lower minimum distance.

\begin{table}[h]
\centering
\scalebox{0.8}{
\begin{tabular}{lccc}
\toprule
\textbf{Variable} & \textbf{7-Variable Model} & \textbf{8-Variable Model} \\
\midrule
$\Omega_{\Lambda_2}$ & - & -2.077437E-08 \\
$\Omega_{\Lambda 3}$ & 0.00233471 & 0.00432285 \\
$h$ & 0.67733813 & 0.67537001 \\
$\omega_b$ & 0.02238564 & 0.02238631 \\
$\omega_{cdm}$ & 0.11984093 & 0.11985634 \\
$A_s$ & 2.094146E-09 & 2.09484E-09 \\
$n_s$ & 0.96599393 & 0.96583363 \\
$\tau_{reio}$ & 0.05295666 & 0.0530449 \\
\midrule
predicted distance & 27.877821 & 24.391195 \\
\bottomrule
\end{tabular}
}
\caption{Optimized feature values for minimum distance in 7- and 8- parameter models}
	\label{table2}
\end{table}
Before proceeding to the robustness analysis with bootstrapping, it is useful to check whether the best-fit values of the finite-temperature parameters are consistent with existing observational bounds from early dark energy. Let us check consistency with the phenomenological early-dark-energy bounds: the tightest current constraint places the EDE fraction at decoupling ($z\sim 1000$--$2000$) below $\sim 0.4$--$0.6\%$ \cite{Gomez-Valent:2021cbe}. At the best-fit values obtained below ($\Omega_{\L_2}\sim-2\times10^{-8}$, $\Omega_{\L_3}\sim 4\times10^{-3}$), the net finite-temperature energy-density fraction across that window is at the $\sim 10^{-4}$ ($\approx 0.01\%$) level -- well beneath the bound, and consistent with the quantum origin of these terms.

\subsubsection{The results with bootstrap resampling}

\begin{table}[h]
\centering
\scalebox{0.78}{
\hspace{-.3in}
\begin{tabular}{lcc}
\toprule
\textbf{Variable} & \textbf{7-Variable Model (Mean [95\% CI])} & \textbf{8-Variable Model (Mean [95\% CI])} \\
\midrule
$\Omega_{\L_2}$ & — & -2.070386E-08 [-2.073444E-08, -2.068444E-08] \\
$\Omega_{\L_3}$ & 2.336606E-03 [2.331111E-03, 2.338889E-03] & 4.282391E-03 [4.2700E-03, 4.304444E-03] \\
$h$ & 0.6773487 [0.6773333, 0.6780] & 0.6753918 [0.6753333, 0.6755] \\
$\omega_b$ & 0.02239074 [0.02238667, 0.02239778] & 0.02239451 [0.02239444, 0.02239444] \\
$\omega_{cdm}$ & 0.1198824 [0.1196667, 0.1198889] & 0.1198889 [0.1198889, 0.1198889] \\
$A_s$ & 2.098202E-09 [2.096667E-09, 2.101111E-09] & 2.099984E-09 [2.098889E-09, 2.103333E-09] \\
$n_s$ & $0.9660145$ [$0.9660$, $0.9665$] & $0.9659412$ [$0.9658889$, $0.9660$] \\
$\tau_{reio}$ & $0.05385517$ [$0.05350$, $0.0548333$] & $0.05422211$ [$0.05400$, $0.05500$] \\
\midrule
predicted distance & $33.30075$ [$33.08834$, $33.6667$] & $33.35167$ [$33.06932$, $33.69121$] \\
\bottomrule
\end{tabular}
}
\caption{Optimized Feature Values and 95\% Confidence Intervals}
\label{table3}
\end{table}

To rigorously assess the variability and robustness of the regression-based optimization process, we incorporate a bootstrap resampling procedure - a widely used non-parametric technique in statistical analysis. This approach allows us to empirically quantify the uncertainty in our best-fit cosmological parameters and predicted distance values, independent of any prior distributional assumptions. 

In this analysis, we perform 1,000 bootstrap resampling iterations for both the 7-parameter and 8-parameter datasets. In each iteration, the training set (comprising 70\% of the full dataset, i.e., 3,500 samples) is resampled with replacement to generate a new training subset of the same size. The quartic polynomial regression model - identified earlier as the best-performing model based on $R^2$ and mean squared error (MSE) - is then refitted using the resampled data. For each refitted model, we identify the parameter configuration that minimizes the predicted distance, effectively simulating the optimization process under perturbed training data conditions. This repeated process yields a distribution of best-fit parameter values and predicted distance outcomes across the 1,000 bootstrap samples.

From these distributions, we compute empirical 95\% confidence intervals (CIs) by extracting the 2.5th and 97.5th percentiles of each parameter’s distribution. These intervals provide a direct, data-driven estimate of the sampling variability inherent in our modeling framework. Unlike Bayesian credible intervals, which require specification of prior distributions, bootstrap-derived CIs rely solely on the observed data and model fit, making them especially suitable for complex, high-dimensional systems where formal likelihood functions may be intractable or computationally expensive.

The results of this analysis, summarized in Table~\ref{table3}, reveal consistently narrow confidence intervals for most cosmological parameters in both the 7- and 8-parameter models. This suggests a high degree of robustness in the regression-based optimization process: the estimated values are relatively stable across random perturbations of the input data. The predicted distance values likewise exhibit limited variability under bootstrap resampling.

Interestingly, the inclusion of the additional parameter $\Omega_{\Lambda_2}$ in the 8-parameter model introduces modest shifts in the distributions of several other parameters - most notably the Hubble parameter $h$, the scalar spectral index $n_s$, and the reionization optical depth $\tau_{reio}$. These shifts are accompanied by slightly broader confidence intervals and increased standard deviation in the predicted distance values, indicating a mild increase in model complexity and sensitivity to input variability. Nevertheless, the confidence intervals for these parameters in both models overlap substantially, suggesting that the two regression models produce statistically consistent outputs.

Taken together, the bootstrap results lend strong support to the generalizability and robustness of our quartic regression models. They confirm that the optimized parameter values obtained via our data-driven procedure are not artifacts of a particular sample realization, but instead reflect stable features of the underlying parameter-distance relationship as encoded in the CLASSy-generated datasets.

\subsection{Comprehensive comparison: $\Lambda$CDM vs. finite-T models}

Having presented the detailed analysis of both the standard $\Lambda$CDM model with spatial curvature and the quantum gravitational extensions incorporating $\Omega_{\Lambda_2}$ and $\Omega_{\Lambda_3}$, we now provide a comprehensive comparative assessment of these three cosmological frameworks. 
\begin{figure}
	\centerline{
		\begin{minipage}[b]{6cm}
			\epsfxsize=6cm
			\epsfbox{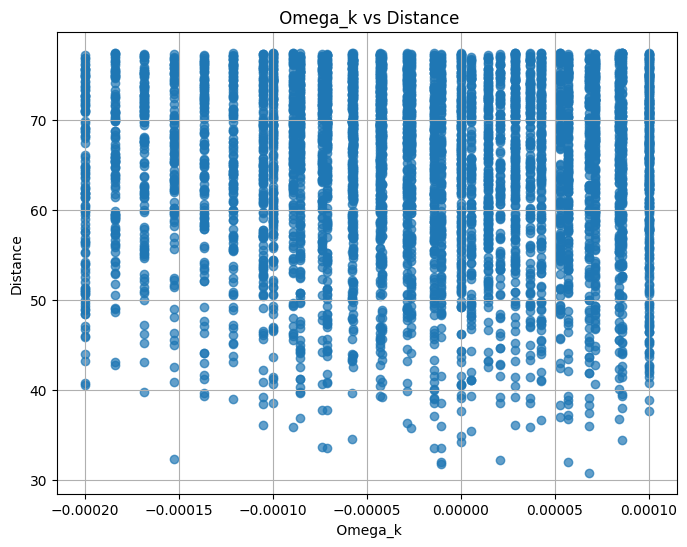}
		\end{minipage}\;\;\;
		\begin{minipage}[b]{6cm}
			\epsfxsize=6cm
			\epsfbox{"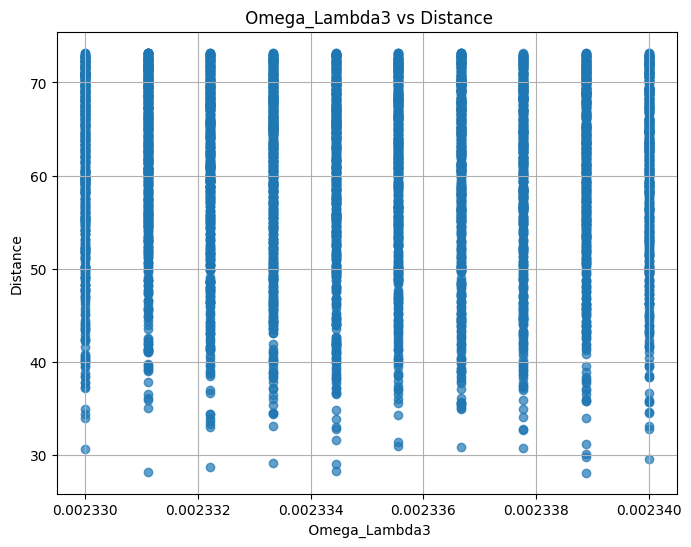"}
		\end{minipage}
		\begin{minipage}[b]{6cm}
			\epsfxsize=6cm
			\epsfbox{"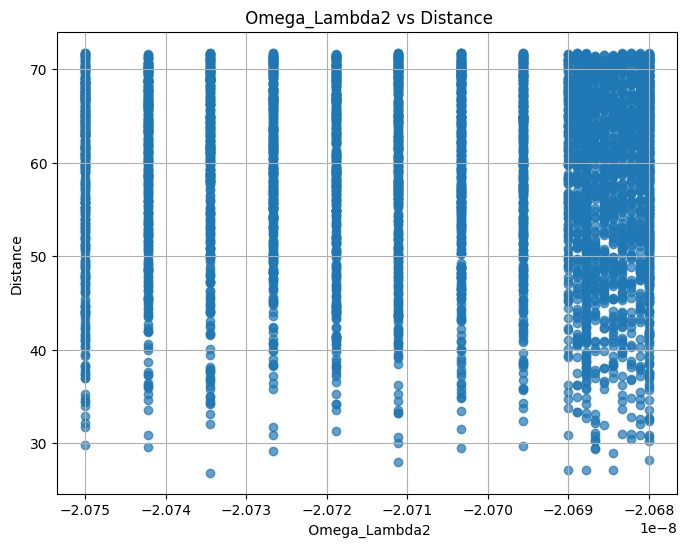"}
		\end{minipage}
	}
\vspace{-.1in}	
\[	\;\;\;\;\hspace{-.0in}\mbox{(a)} \hspace{2.4in}\mbox{(b)} \hspace{2.2in}\mbox{(c)}\]
\vspace{-.3in}
	\caption{Scatter plots illustrating the dependence of the Euclidean distance to the Planck 2018 power spectrum on selected cosmological parameters across the three model frameworks considered. (a) Distance versus spatial curvature parameter $\Omega_K$ for the $\Lambda$CDM model, showing a relatively broad spread of distances and higher minimum values. (b) Distance versus quantum gravitational parameter $\Omega_{\Lambda_3}$ for the 7-parameter quantum-corrected model, demonstrating an overall downward shift in achievable distances compared with the standard model. (c) Distance versus $\Omega_{\Lambda_2}$ for the 8-parameter quantum-corrected model including both $\Omega_{\Lambda_2}$ and $\Omega_{\Lambda_3}$, highlighting further reduction in minimum distance and improved agreement with observational constraints. Together, the panels illustrate the systematic improvement in model-data concordance obtained by incorporating finite-temperature quantum gravitational corrections.}
	\label{disditri}
\end{figure}
The most striking difference among the three models emerges in their distance distributions and achievable minima; see Fig. \ref{disditri}. The standard $\Lambda$CDM model exhibits the highest minimum distance (30.78) and largest mean distance (64.62 $\pm$ 9.64). In contrast, the quantum gravitational models demonstrate systematically superior agreement with observational constraints: the 7-parameter model (with $\Omega_{\Lambda_3}$ only) achieves a minimum distance of 28.12 and mean of 55.75 $\pm$ 8.19, while the 8-parameter model (with both $\Omega_{\Lambda_2}$ and $\Omega_{\Lambda_3}$) reaches an even lower minimum of 26.83 with mean 58.80 $\pm$ 9.70.

Remarkably, when optimized using regression-based techniques, the quantum models predict even lower distances: 27.88 (7-param) and 24.39 (8-param), which upon CLASS verification yield actual distances of 26.03 and 24.95, respectively. This represents approximately 15-19\% improvement over the standard $\Lambda$CDM optimized minimum of 29.56, indicating that the quantum gravitational parameterizations reproduce the fine structure of the reference spectrum more closely than spatial curvature modifications alone. We stress that this is a statement about fit to the fixed Planck reference curve, not a likelihood-based preference of the data for the extended model (see Table 6).

\begin{table}[h]
\centering
\small
\begin{tabular}{lccc}
\toprule
\textbf{Metric} & \textbf{$\Lambda$CDM 7-param} & \textbf{finite-T 7-param} & \textbf{finite-T 8-param} \\
\midrule
Min Distance (data) & 30.78 & 28.12 & 26.83 \\
Mean Distance & 64.62 & 55.75 & 58.80 \\
Distance Std Dev & 9.64 & 8.19 & 9.70 \\
Predicted Min & 29.56 & 27.88 & 24.39 \\
Verified Min & --- & 26.03 & 24.95 \\
\bottomrule
\end{tabular}
\caption{Distance metric comparison}
\label{tab:distance_comparison}
\end{table}

\subsubsection{Regression performance and model complexity}

All three models adopt degree-4 polynomial regression as optimal based on AIC/BIC criteria. However, their predictive performance differs substantially:
\begin{table}[H]
\centering
\small
\begin{tabular}{lcccccc}
\toprule
\textbf{Model} & \textbf{$R^2_{\rm train}$} & \textbf{$R^2_{\rm test}$} & \textbf{MSE$_{\rm test}$} & \textbf{AIC} & \textbf{BIC} \\
\midrule
$\Lambda$CDM 7-parameter & 0.9924 & 0.9912 & 0.8281 & 471.4 & 2091.0 \\
Finite-T 7-parameter & 0.9990 & 0.9988 & 0.0795 & $-3138.5$ & $-1385.1$ \\
Finite-T 8-parameter & 0.9992 & 0.9988 & 0.1096 & $-2325.9$ & 304.1 \\
\bottomrule
\end{tabular}
\caption{Regression performance metrics}
\label{tab:regression_comparison}
\end{table}
\ni The quantum gravitational models achieve closer fits, with test MSE values approximately 7-10 times lower than standard $\Lambda$CDM (0.08-0.11 vs. 0.83). The vastly improved AIC values ($-3138$ and $-2326$ vs. $+471$) indicate that quantum parameterizations provide substantially better model-data concordance even after penalizing for complexity. The BIC values follow similar trends, though the 8-parameter model's positive BIC (304.1) reflects the substantial complexity penalty associated with the additional parameter. Cross-validation results reinforce these conclusions. The standard $\Lambda$CDM model shows Test MSE = $0.858 \pm 0.068$ across 10 folds, whereas the quantum models show lower values.

For convenience, we tabulate the \emph{differences} of the information criteria relative to $\Lambda$CDM (taking the baseline values $\mathrm{AIC}_{\Lambda\mathrm{CDM}}=471.4$, $\mathrm{BIC}_{\Lambda\mathrm{CDM}}=2091.0$ of Table~\ref{tab:regression_comparison}):
\begin{center}
\begin{tabular}{lrrrr}
\toprule
Model & AIC & $\Delta$AIC & BIC & $\Delta$BIC \\
\midrule
$\Lambda$CDM (7-param, $\Omega_K$) & $471.4$ & --- & $2091.0$ & --- \\
finite-T 7-param ($\Omega_{\L_3}$) & $-3138.5$ & $-3609.9$ & $-1385.1$ & $-3476.1$ \\
finite-T 8-param ($\Omega_{\L_2},\Omega_{\L_3}$) & $-2325.9$ & $-2797.3$ & $304.1$ & $-1786.9$ \\
\bottomrule
\end{tabular}
\end{center}
\ni where $\Delta\mathrm{AIC}\equiv\mathrm{AIC}_{T}-\mathrm{AIC}_{\Lambda\mathrm{CDM}}$ (and similarly for BIC), so that a negative value favors the finite-T model. 

We must, however, be explicit about what these criteria do and do not establish, both for accuracy and to forestall a misreading. The AIC and BIC values above are computed from the \emph{surrogate} quartic regression -- the polynomial mapping the cosmological parameters to the Euclidean distance from the Planck curve. They quantify how well the finite-T parameterization allows that parameter-to-distance surface to be fit, and by how much the fit improves relative to the $\Lambda$CDM surrogate; in that role they are genuine and informative. They are \emph{not} computed from a cosmological likelihood, and therefore should not be read directly on the Jeffreys evidence scale as a Bayesian statement that the Planck data ``prefer'' the finite-T model over $\Lambda$CDM at a positive/strong/very-strong significance level. Such a statement requires a proper likelihood-based inference, which is the subject of a companion Bayesian analysis (see below and the conclusion): a Cobaya/CLASS Markov-chain treatment (modified CLASS with $\Omega_{\L_2}$ and $\Omega_{\L_3}$, Planck 2018 likelihood) from which the deviance information criterion (DIC) and its Jeffreys-scale reading are naturally obtained. That analysis finds the finite-T parameters to be only weakly constrained by Planck alone -- in particular $\Omega_{\L_3}$ has a broad, one-sided posterior consistent with zero, because its leading effect on the background expansion is largely absorbable into a compensating shift of $H_0$ -- so the formal likelihood-based evidence is not at the ``strong'' level, and we make no such claim here. What the present surrogate analysis robustly establishes is the more modest statement that, within the observationally allowed room, the finite-T parameters improve the fit to the fine structure of the power spectrum, and that this improvement is not merely an artifact of added polynomial flexibility (the 7-parameter $\Lambda$CDM and finite-T surrogates share identical dimensionality after polynomial expansion, yet differ markedly in fit quality). The quantitative interpretation of $\Delta$AIC/$\Delta$BIC on the Jeffreys scale can be found in the statistical glossary (Appendix~A).

\subsubsection{Parameter degeneracies and correlations}

The introduction of the finite-T parameters induces notable shifts in other cosmological parameters relative to the standard $\Lambda$CDM baseline. The Hubble parameter $h$ decreases slightly compared with the standard model, which yields $h = 0.679$ (optimized) or $0.6796$ (bootstrap mean), whereas the quantum models prefer lower values of $0.6773$ for the 7-parameter model and $0.6754$ for the 8-parameter model, with the addition of $\Omega_{\Lambda_2}$ systematically shifting $h$ downward. The spectral index $n_s$ remains remarkably stable across all frameworks, taking values of $0.9659$ for $\Lambda$CDM, $0.9660$ for the 7-parameter quantum model, and $0.9658$ for the 8-parameter quantum model, suggesting that $n_s$ is well constrained regardless of the dark energy parameterization. The reionization optical depth $\tau_{\rm reio}$ shows modest increases from $0.05089$ in the optimized $\Lambda$CDM model to $0.05296$ and $0.05304$ in the 7- and 8-parameter quantum models, respectively, indicating a preference for slightly higher reionization optical depths in the quantum frameworks. Meanwhile, the primordial amplitude $A_s$ remains highly consistent across all models, with values of $2.086 \times 10^{-9}$ for $\Lambda$CDM and $2.094 \times 10^{-9}$ for both quantum models, demonstrating strong robustness to these model extensions.
Bootstrap analysis reveals overlapping 95\% confidence intervals for most parameters between the quantum models (Table \ref{table3}) and the standard $\Lambda$CDM model (Table \ref{tab:lcdm_bootstrap}), indicating statistical consistency despite the improved fit quality.

\subsubsection{Computational efficiency, stability, and implications for model selection}

Learning curve analysis reveals that the quantum 7-parameter model stabilizes around 1,000 training samples, while the 8-parameter model requires approximately 2,500 samples due to increased complexity from $\Omega_{\Lambda_2}$. In comparison, the standard $\Lambda$CDM model stabilizes at intermediate complexity, though this analysis was not explicitly documented in the provided results. Cross-validation stability metrics show that standard $\Lambda$CDM exhibits minimum predicted distance variability of $34.71 \pm 1.38$ across folds. 
The comprehensive comparison yields several key conclusions: the finite-T parameterizations demonstrate a clear advantage over standard $\Lambda$CDM \emph{in surrogate fit quality}, achieving 7-10$\times$ lower MSE and dramatically improved information criteria, AIC and BIC, thereby indicating a substantially better fit of the surrogate to the parameter--distance surface. We reiterate that these are surrogate-regression diagnostics against the fixed Planck reference spectrum, not likelihood-based evidence that the data prefer the extended model; the latter question is deferred to the companion Bayesian analysis, where the parameters are found to be weakly constrained. This improvement is further reflected in the minimum-distance analysis, where the 15-19\% reduction in achievable minimum distances (24.95 compared with 29.56) indicates that quantum corrections reproduce the fine structure of the reference spectrum more closely than spatial curvature modifications alone. Despite these improved fits, the resulting cosmological parameters remain broadly consistent with those of standard $\Lambda$CDM within bootstrap confidence intervals, avoiding pathological or unphysical parameter values and preserving overall parameter stability. From a theoretical standpoint, the negative $\Omega_{\Lambda_2}$ coefficient and modest-magnitude quantum corrections ($\Omega_{\Lambda_3} \sim 10^{-3}$) are physically plausible and align with expectations from quantum field theory in curved spacetime; see the analysis below. The 8-parameter model yields the lowest minimum distance, its positive BIC reflects non-trivial penalties associated with increased model complexity.

From the standpoint of the surrogate analysis, the finite-T parameterizations yield a markedly better fit to the parameter--distance surface than standard $\Lambda$CDM with spatial curvature. This improvement cannot be attributed solely to additional parameters, as the 7-parameter frameworks (standard vs. quantum) have identical formal dimensionality after polynomial expansion yet exhibit vastly different predictive performance. As emphasized above, this is a statement about surrogate fit quality; the likelihood-based question of whether the data \emph{prefer} the extended model is deferred to the companion Bayesian analysis, where the parameters are found to be weakly constrained.
However, several caveats warrant emphasis. First, this analysis constitutes a model-building exercise using CLASS-generated synthetic data rather than direct observational inference. The minimum distance metric, while useful for optimization, does not directly correspond to standard cosmological likelihood functions. Second, the bootstrap confidence intervals, though empirically robust, do not constitute full Bayesian posterior distributions and thus cannot formally assess parameter degeneracies through covariance structure. 

Nevertheless, the consistent pattern of improved surrogate performance across multiple evaluation criteria (MSE, $R^2$, AIC, BIC, cross-validation, minimum achievable distances) indicates that quantum gravitational parameterizations warrant serious consideration as extensions to standard $\Lambda$CDM cosmology, and motivates the dedicated likelihood-based study reported separately. That the improvements are consistent across criteria, and are not removed by the polynomial-degree penalty, indicates they reflect genuine structure in the parameter--distance relation rather than mere overfitting of the surrogate;\footnote{Crucially, for every model the training and testing $R^2$ differ by at most a few hundredths of a percentage point ($\L$CDM: 99.25\% vs. 99.07\%; 7-parameter: 99.90\% vs. 99.88\%; 8-parameter: 99.92\% vs. 99.88\%). Since the standard diagnostic for overfitting is a large train-test gap, these near-identical values - together with the plateaued learning curves - confirm that the surrogate generalizes and is not overfitting.} whether they correspond to a statistically preferred cosmological model is the separate question addressed by the upcoming Bayesian analysis. (See the related comments in the conclusion.)

\subsection{(Un)naturalness of the order of $\Omega_{\L_2}$ \label{OL2na}}

We address the theoretical interpretation of the magnitude of the finite-temperature parameter $\Omega_{\Lambda_2}$ inferred from the numerical analysis. Although its fitted value appears smaller than naive expectations based on earlier renormalization arguments, we emphasize that $\Omega_{\Lambda_2}$ is treated phenomenologically in the present work. Its magnitude is sensitive to the choice of renormalization scheme and to the treatment of mass-dependent terms in the finite-temperature expansion. We argue that the apparent smallness of $\Omega_{\Lambda_2}$ does not necessarily signal unnatural fine-tuning, but may instead reflect the limitations of the lowest-order approximation or the need for an alternative renormalization prescription. These issues are explored in detail below, and possible resolutions are outlined.

In the numerical fits (sections~2--3.3), the $T^4$ and $T^2$ terms are the standard leading terms of the high-temperature expansion \rf{JmT}, evaluated with the ordinary Standard Model particle content, and $\Omega_{\L_2},\Omega_{\L_3}$ are treated \emph{phenomenologically} -- their magnitudes are fit to the data, not fixed by any choice of mass. The prescription of setting the renormalized mass on the order of the temperature enters \emph{only} in the present subsection, whose separate and more speculative aim is to ask whether the \emph{magnitude} of $\Omega_{\L_2}\sim 10^{-8}$ can be understood as natural. (This scheme is slightly different from the one \cite{Park:2021ohu,Park:2021vro} adopted to alleviate the CC problem, but can be viewed as its variation.) Consequently, the $T^4$/$T^2$ phenomenology of the preceding sections stands independently of one's stance on the CC problem or on any particular renormalization scheme. We note in passing that the perspective adopted here -- that the CC problem may be a matter of renormalization-scheme choice rather than a genuine physical problem -- is shared by other recent work, e.g.\ \cite{Ageeva:2024qie}.

As mentioned earlier, the crux of our approach lies in the introduction of $\Omega_{\L_2}$ (and $\Omega_{\L_3}$). The analysis in section 3.2 shows that $\Omega_{\L_2}$ is on the order of $10^{-8}$. We view these parameters as phenomenological. However, below, we will discuss the (un)naturalness of the order of $\Omega_{\L_2}$, specifically its value of $10^{-8}$, by introducing a different renormalization scheme (a different one than that leading to \rf{mainres}) that is better suited for this purpose.

In anticipation of the analysis below, let us bring home the distinction between the zero-temperature contributions and the finite-temperature contributions. We employ a quantum field theoretic matter system, as opposed to a hydrodynamic one for the CLASS-related analysis in section 3.2, for the present discussion. For the CC problem, the role of the temperature-dependent terms was more highlighted in \cite{Park:2021ohu}, \cite{Park:2021vro}. This difference is due to the slight change of the renormalization schemes. Let us recall the renormalization of the cosmological constant proposed in \cite{Park:2021ohu} \cite{Park:2021vro}, and \cite{Park:2024kfn}. It was demonstrated that the one-loop correction
is two or three orders of magnitude smaller than the observed value. The essential idea can be explained by considering a scalar system: 
\bea
S_{\text{ren}}
&=& -\int d^4x \Bigg[ \frac{1}{2}  \partial_\mu \phi  \partial^\mu \phi + \frac{1}{2} m^2 \phi^2 + \frac{\lambda}{4!}  \phi^4 \Bigg].
\label{arnold}
\eea
The cosmological constant (CC) value depends, of course, on whether one adopts the complete-square form or the version without the constant term. (Whether to use the complete-square form or the traditional one is not central to the CC problem.) 
For the onshell potential value, one gets \cite{Arnold:1992rz}\cite{Laine:2016hma} 
\bea 
&&V =-\frac{1}{90}  \pi ^2 T^4+\cdots  \la{mainres}
\eea
where the ellipsis includes terms that depend on the mass of the scalar. Let us choose the renormalized mass value appropriately to deal with the CC problem: set the renormalized mass of the scalar particle so that it is on the order of the temperature.\footnote{We emphasize that this is the imposition of a \emph{renormalization condition}, not the postulation of a physical particle of mass $\sim T$: the renormalized mass is a scheme-dependent parameter, kept free throughout the calculation and fixed only by a chosen condition, and is distinct from the pole (physical) mass. In the present scheme the renormalized mass is fixed first, and matching to the physical mass then determines the finite counterterm; this is a legitimate finite renormalization whose consistency is assured by bare perturbation theory. The large freedom exploited here is nothing but the familiar arbitrariness of the renormalization procedure -- there are infinitely many schemes differing by finite renormalizations, and one may change scheme when a choice convenient for one problem becomes inconvenient for another. In the full SM-coupled-to-gravity calculation \cite{Park:2024kfn} every field (Higgs, massive gauge bosons, fermions, graviton) contributes its own $J(m,T)$, the scalar of \rf{Voneloop}--\rf{JmT} serving only as the representative field for the illustrative derivation.} (See \cite{Moreno-Pulido:2020anb} and references therein for earlier works where a CC resolution proposal involves a novel renormalization scheme.\footnote{
This is the running vacuum model (RVM)---a recent comprehensive review of which can be found in \cite{SolaPeracaula:2026pgi}. 
As shown in \cite{SolaPeracaula:2026trz}, the RVM is closely related to the decay of the de Sitter vacuum into a radiation-dominated FLRW epoch, 
with both scenarios driven by quantum field theory effects that renormalize the vacuum energy density \(\rho_{\text{vac}}\) via adiabatic prescriptions. A potential connection with temperature-dependent effects in the early universe was also noted. This mechanism, known as RVM-inflation, offers a unified QFT description of both inflation and dark energy. 
Furthermore, as demonstrated in \cite{SolaPeracaula:2026lyk}, the composite version of the RVM, the so-called \(\Lambda\)XCDM, naturally accounts for the phantom-divide crossing 
observed by DESI near \(z \simeq 0.4\), with the cosmon component behaving as phantom matter  
that satisfies the strong energy condition.}) At the same time let it be sufficiently smaller than the temperature so that their (i.e., mass-dependent) contributions to the CC can be disregarded. With this renormalization scheme, the leading term in the potential is $\sim T^4$, as indicated above. Evaluated at the present temperature of the Universe, this implies that there is a order $10^{4-5}$ gap between the value above and the observed value of the CC, a (potentially) big improvement from the gap noted in the CC problem. As anticipated in \cite{Park:2021ohu}, once one considers the Standard Model, its particle content should further reduce the gap. To see this, note that a fermionic field contributes
\bea
\fr78 \fr{\pi^2}{90}T^4+\cdots
\eea
to the potential. By going over the contributions of the SM particles, which provides an additional factor of around 120, the CC order discrepancy is now reduced from $10^{60}$ to $10^{2-3}$.\footnote{The details of renormalization analysis can be quite subtle. For the zero-T component, the bosonic and fermionic contributions tend to cancel each other out due to their opposite signs. However, for the $T^4$ term, the situation is different: the one-loop bosonic and fermionic contributions carry a negative sign, so the fermionic contributions add to the bosonic ones. (In $\overline{\mbox{MS}}$ scheme this implies that the finite part of the CC has a negative sign.)
Given that there are more fermionic degrees of freedom than bosonic ones in the SM (coupled with gravity), the relative magnitudes of these contributions must be carefully balanced by adjusting the renormalized masses of the bosonic and fermionic fields to ensure the correct sign for the overall CC. This balancing act can be helped by adjusting, if necessary, the finite parts of the renormalization parameters, often denoted as $c$'s in renormalization theory. These $c$'s play a crucial role in the proposed resolution of the CC problem. \la{msign}} This observation was based on the one-loop correction, which has been shown to be two or three orders of magnitude smaller than the observed value, as just stated. 
However, it is possible to introduce a classical contribution to the cosmological constant (CC), and it is standard practice in renormalization to include such a classical term, the renormalized CC, $\L_{ren}$, with the corresponding expression added to the density parameter. With this approach, the unnaturalness of adding or subtracting extremely large and small numbers is no longer present \cite{Park:2021ohu,Park:2021vro,Park:2024kfn}. (This approach, however, comes with a trade-off, as reviewed below.)

\vspace{.3in}

Let us now consider the naturalness of the order of magnitude of $\Omega_{\L_2}$, which has been found to be on the order of $10^{-8}$. The solution to the CC problem proposed in \cite{Park:2021ohu, Park:2021vro, Park:2024kfn}, which was briefly reviewed above, suggests that $\Omega_{\L_2}$ should be related to $\Omega_{\L_1}$ through a factor of $10^{-(2 \sim 3)}$, i.e., $\Omega_{\L_2} \sim \Omega_{\L_1} \times 10^{-(2\sim 3)}$. Given that $\Omega_{\L_1} \simeq 0.7$, the natural expectation for the order of $\Omega_{\L_2}$ is $\Omega_{\L_2} \simeq 10^{-(3\sim 4)}$. However, the numerical analysis carried out in the present study yields a value of $\Omega_{\L_2} \sim 10^{-8}$, which is several orders of magnitude smaller than the expected range of $10^{-(3\sim 4)}$.

This discrepancy, where the order of $10^{-8}$ seems much smaller than the expected $10^{-(3\sim 4)}$, raises the question of whether this result is unnatural, and if so, whether it can be explained. It would indeed be beneficial to explore this gap further and attempt to provide at least some level of explanation for it. However, before delving into a more detailed analysis, let us briefly recall the nature of the proposal in \cite{Park:2021ohu, Park:2021vro, Park:2024kfn} and the challenges it presents.

The solution of the CC problem suggested in \cite{Park:2021ohu, Park:2021vro, Park:2024kfn} and reviewed above is not without its "cost." This cost, however, is not unique to the proposal at hand, but rather reflects a more general phenomenon inherent in renormalization schemes. The approach used in those works involves an appropriate renormalization scheme, and as such, it is not free from a certain degree of unnaturalness when applied to other problems. As well known, physical quantities must be independent of the renormalization schemes used. There are infinitely many renormalization schemes that differ by finite renormalization. It is possible, and often necessary, to change the renormalization scheme when an initial choice made for one problem becomes inconvenient for another. 

In the case of the CC problem, this unnaturalness, which is nothing but the well-known fine-tuning, arises specifically due to the onshell renormalization scheme. This scheme uses the physical value of certain parameters, such as the Higgs mass, directly, which leads to a contribution to the CC that is much larger than the observed value. This prompts an important question: if using the physical mass in the onshell scheme leads to this issue, why not consider an alternative renormalization scheme that could potentially avoid or mitigate the problem? See \cite{Ageeva:2024qie} (and references therein) for a recent work with similar ideas.)

In this spirit, the proposal in \cite{Park:2021ohu, Park:2021vro} suggests an alternative approach: rather than using the physical masses directly, one could adopt renormalized mass values that are sufficiently small to avoid causing fine-tuning issues. More specifically, it is suggested that mass values on the order of the temperature might be more appropriate. This would ideally lead to a solution that is less prone to the unnaturalness arising from the onshell renormalization scheme.

The current situation, despite appearing somewhat unnatural with $\Omega_{\L_2} \sim 10^{-8}$, may not necessarily represent an insurmountable obstacle. Instead, it could reflect a situation where the issue might be resolved by adopting a different renormalization scheme, a variation of the one adopted to avoid the CC problem. The key feature of this alternative scheme is that the masses are adjusted in such a way that the mass-dependent terms in the potential of eq. \rf{mainres}, which were neglected in the previous renormalization scheme applied to the CC problem, are kept small but not negligibly so. (Recall that in the renormalization scheme that led to \rf{mainres}, the mass was neglected compared with T.) This should lead to a slower convergence in the large-temperature expansion since the order of $\frac{m}{T}$,  where $m$ denotes the value of the renormalized mass of, say, a Higgs particle, is larger than before. This slower  convergence\footnote{ Slow convergence {\em compared with the standard renormalization scheme} was also expected in the proposed CC resolution in \cite{Park:2021ohu, Park:2021vro}.} could, in turn, suggest that the renormalized cosmological constant $\L_{ren}$ should be set to a value smaller than the observed cosmological constant $\L_{obs}$, by several orders of magnitude. This contrasts with the approach used in the CC problem resolution, where the renormalized and observed values were taken to be on the same order.

Such an adjustment would allow for a potential compensation of the deficit through higher-order terms that slowly converge. As a result, the observed value of $\Omega_{\L_2} \sim 10^{-8}$ might not appear as unnatural after all. It is conceivable that higher-order terms in the expansion could shift the one-loop cosmological constant to a value closer to the observed value, thus bridging the gap between the theoretical estimate and the actual observed data.

The situation could also unfold differently on the numerical side. There is a possibility that by repeating the numerical analysis in section 3, and including higher-order terms in the calculation, the optimal value of $\Omega_{\L_2}$ could increase from $\sim 10^{-8}$. In this case, the inclusion of these terms might lead to a larger value for $\Omega_{\L_2}$, which could help reconcile the numerical result with expectations based on the proposed renormalization scheme. This speculation suggests that the value of $\Omega_{\L_2}$ might not be fixed and could evolve as higher-order corrections are included, offering a potential resolution to the discrepancy.

In conclusion, while the order $10^{-8}$ might initially seem unnatural, it is possible that the issue could be addressed through the use of an alternative renormalization scheme that adjusts the mass parameters in a way that makes the solution more natural, or that higher-order terms could lead to a more reasonable outcome. Further analysis is required to clarify the situation, and we will address these possibilities in the conclusion.

\section{Conclusion}

In a flat background, it is a textbook result that finite-temperature effects generate vacuum energy terms. It is natural to expect that this result will be generalized to a curved spacetime, leading to cosmological ``constant'' terms. This expectation has been verified \cite{Park:2024kfn} within the framework of the foliation-based quantization of gravity \cite{Park:2014tia,Park:2015qxa,Park:2015xoa}. The quantum corrections modify, among other things, the cosmological constant term. This work demonstrates the impact of the shifted cosmological constant on the fine structure of the power spectrum and its implications for the broader set of cosmological parameters. Our analysis suggests that perturbative quantum gravitational effects must be accounted for in periods around recombination, as these effects are not negligible. More specifically, through the introduction of additional parameters, such as $\Omega_{\Lambda_2}$ and $\Omega_{\Lambda_3}$, we have extended the standard $\Lambda$CDM model, offering a more sophisticated approach to fitting cosmological data. By applying the CLASS and utilizing both brute-force scans and machine learning regression techniques, we have demonstrated how the temperature-corrected cosmological parameters can lead to a more accurate fit to the Planck 2018 data. The examination of the 7- and 8- parameter models based on quartic regression has revealed that the presence of small nonzero values of $\Omega_{\Lambda_2}$ and $\Omega_{\Lambda_3}$ measurably improves the fit to the fine structure of the spectrum and has shown promise in refining the model while preserving its generalizability. We reiterate that this is an improvement in surrogate fit quality against the fixed Planck reference: their measurable effect is confined to the shape residual against the Planck curve, where the achievable minimum distance falls by roughly 15–19\%. The question of whether current data statistically prefer the extended model is addressed by the companion Bayesian analysis, where the finite-T parameters are only weakly constrained.

To complement the rigid selection criteria, we employed machine-learning validation techniques, including artificial neural networks and leave-one-out cross-validation, as discussed in~\cite{Hatefi:2025lqe}. Both the results reported in that work and those obtained here consistently yield high values of the coefficient of determination $R^2$ together with low mean squared errors across validation folds. This behavior indicates that the improved fit does not arise from overfitting, but instead reflects genuine structure in the parameter--distance relation. Sensitivity analyses further show that the inclusion of $\Omega_{\Lambda_2}$ plays a central role in achieving this fit quality. Its removal leads to a more pronounced degradation in performance than the exclusion of some of the standard $\Lambda$CDM parameters, suggesting that the finite-temperature contributions help alleviate a structural misspecification in the standard model, wherein cosmological parameters partially compensate for missing early-Universe physics. The results presented in Table~3 and Fig.~7 in \cite{Hatefi:2025lqe} reinforce this interpretation: while the overall large-scale behavior of the curve is governed primarily by the standard parameters, $\Omega_{\Lambda_2}$ and $\Omega_{\Lambda_3}$ are crucial for determining its fine structure. In particular, when attention is restricted to curves that closely reproduce the \textit{Planck} 2018 results, these finite-temperature parameters become essential for achieving an accurate match.

\vspace{.2in}

While the results presented here reinforce the utility of finite-$T$ QFT effects in cosmological studies, they also open up avenues for future research. The natural complement to the present surrogate analysis is a full Bayesian parameter estimation using the Planck 2018 likelihood, which provides the likelihood-based cross-check that the surrogate diagnostics cannot supply and quantifies the parameter uncertainties systematically. This more standard, and computationally intensive, approach is the subject of the companion study in preparation \cite{inPrep}; as summarized in Section~3.3, its preliminary indication is that the finite-T parameters are only weakly constrained by Planck alone, consistent with the near-invariance of the standard parameters found here. A dedicated account, including the DIC and its Jeffreys-scale reading, is deferred to that work.

One potential improvement can be seen from the residual plots in Figs.~\ref{polyeq7} and \ref{polyeq8}, which display certain patterns for both models, particularly for the 8-variable model. While the residuals remain centered around zero, the spread indicates minor deviations from ideal predictions, especially for higher distance values. These residual patterns suggest that while the models are highly accurate, there is room for improvement in capturing some higher-order non-linearities or interactions present in the data. Various different machine learning techniques, including artificial neural networks, have been and are currently being tried. So far, the quartic regression and artificial neural network methods turn out to be best-performing. The result of the artificial neural network \cite{Hatefi:2025lqe} shares the same qualitative features as those obtained here. Some complementary aspects, such as feature importance, are reported therein.

The higher-order terms may well play important roles - an expectation motivated by the slow convergence observed in previous studies of the cosmological constant problem \cite{Park:2021ohu,Park:2021vro}. Refining the current approach to incorporate these higher-order corrections, as well as extending machine learning methods for enhanced predictive accuracy, will thus be important next steps. In particular, it will be worth including an additional parameter associated with the linear temperature term, $\sim \frac{1}{a}$, and examining its impact on the rest of the parameters.\footnote{The ellipsis in eq.~\rf{tdcc} includes both higher-order loop corrections and high-order terms in the $\fr{m}{T}$ expansion. Analyzing the former terms would require significantly greater effort.} While our present numerical study indicates that leading-order finite-temperature effects alone do not resolve the Hubble tension, it remains to be seen whether this further extended model could potentially ameliorate it. In contrast, see, e.g., \cite{SolaPeracaula:2021gxi} for a study in which a novel renormalization scheme achieves a similar goal. Given the exponential growth in the complexity of the parameter space, addressing these higher-order effects and thoroughly exploring their implications - especially with more complete Bayesian methods - will likely require supercomputing resources. We will report progress on these issues elsewhere.

Beyond the phenomenology, we wish to record a methodological point that we regard as a theoretical outcome of this line of work, independent of whether the finite-temperature parameters are ever detected. Both the cosmological-constant problem and the smallness of $\Omega_{\L_2}$ illustrate the same lesson: a magnitude that appears unnatural in a given renormalization scheme need not be unnatural per se, because judgments of naturalness are themselves scheme-dependent. The onshell scheme, natural and convenient for many Standard Model observables and favored by the fast convergence of its perturbation series, carries no privileged claim to naturalness when it is turned on the vacuum energy; recognizing this dissolves the apparent fine-tuning of the CC rather than requiring new physics to explain it \cite{Park:2021ohu,Park:2021vro}. The physically sound response to an apparent fine-tuning is therefore not to accept it as a genuine problem, but to ask whether a different, equally legitimate scheme - related to the first by a finite renormalization - removes it. The same view has been reached independently, from a quite different starting point, by Ageeva, Petrov and Shaposhnikov \cite{Ageeva:2024qie}, who argue that naturalness and fine-tuning problems can be artifacts of the conventional divergent-graph-plus-multiplicative-renormalization formalism rather than genuine physical challenges. The order of $\Omega_{\L_2}$ obtained here is, from this standpoint, a second instance of the same phenomenon that underlies the CC problem. We stress that the specific proposal - that a variational version of the scheme of \cite{Park:2021ohu,Park:2021vro} accounts for $\Omega_{\L_2}\sim 10^{-8}$ - remains speculative and subject to the further analysis noted in section \ref{OL2na}; the underlying principle, that naturalness is scheme-dependent and that the two problems benefit from the same reappraisal, we state with confidence.

\vspace*{.5in}

\ni {\bf Acknowledgements}\\
\vspace*{-.1in}

\ni This project is funded in part by a State of Arkansas Department of Higher Education HIRED Workforce Development Grant, Award No. AWD-105943 \cite{HIRED}. This work used Anvil at Purdue University through allocation
PHY240277 from the Advanced Cyberinfrastructure Coordination
Ecosystem: Services \& Support (ACCESS) program~\cite{ACCESS},
which is supported by U.S. National Science Foundation grants
\#2138259, \#2138286, \#2138307, \#2137603, and \#2138296.

\newpage
\appendix



\renewcommand{\theequation}{A.\arabic{equation}}
\setcounter{equation}{0}

\section{Statistical glossary}

\begin{description}
 \item[\textbf{Regression}:] A statistical technique used to model and analyze the relationship between a dependent variable and one or more independent variables (also called features or predictors). The goal is to estimate how changes in the independent variables influence the dependent variable. Common types include linear regression, where the relationship is modeled as a straight line, and polynomial regression, where higher-degree terms allow modeling of non-linear patterns.

  \item[\textbf{$R^2$ (Coefficient of determination)}:] A statistical measure that represents the proportion of variance in the dependent variable that is predictable from the independent variables. It ranges from 0 to 1, with values closer to 1 indicating better model performance.

  \item[\textbf{MSE (Mean squared error)}:] A measure of the average squared difference between the observed actual outcomes and the outcomes predicted by the model. It is given by
  \[
  \text{MSE} = \frac{1}{n} \sum_{i=1}^{n} (y_i - \hat{y}_i)^2,
  \]
  where $y_i$ is the actual value, $\hat{y}_i$ is the predicted value, and $n$ is the number of observations.

 \item[\textbf{Standardization}:] A preprocessing technique used to scale features so they have a mean of 0 and a standard deviation of 1. This ensures that features contribute equally to the model, especially important for distance-based and regularized methods.

  \item[\textbf{Residuals}:] The difference between observed values and the values predicted by a regression model. For a given observation, the residual is $e_i = y_i - \hat{y}_i$. Residuals are used to diagnose the fit of a regression model.

    \item[\textbf{Learning curve}:] A plot showing how a model's performance (typically accuracy or error) on the training and validation datasets changes as the training set size increases. It helps assess whether adding more data improves model performance.

  \item[\textbf{Overfitting}:] A modeling error that occurs when a regression model is too complex and captures noise rather than the underlying pattern in the data. It usually results in high training accuracy but poor generalization to new data.

\item[\textbf{Cross-validation}:] A statistical method used to estimate the performance of a model on an independent dataset. In $k$-fold cross-validation, the dataset is divided into $k$ subsets; the model is trained on $k-1$ of them and tested on the remaining subset. This process is repeated $k$ times, each time with a different test set.

\item[\textbf{ CI (Confidence interval)}:] 
A confidence interval is a range of values computed from sample data to estimate an unknown population parameter. A $100(1-\alpha)\%$ confidence interval is constructed so that, over repeated sampling, the interval contains the true parameter in approximately $100(1-\alpha)\%$ of cases.

\item[\textbf{Bootstrapping}:] 
A resampling technique used to estimate the statistical uncertainty of a quantity by repeatedly drawing samples (with replacement) from the original dataset. For each resampled dataset, the statistic of interest is recomputed, producing an empirical distribution from which standard errors, confidence intervals, and bias estimates can be obtained. Bootstrapping is particularly useful when analytic error estimates are difficult to derive.

\item[\textbf{AIC (Akaike information criterion)}:] 
A measure used for model comparison that balances goodness of fit against model complexity. It is defined as
\[
\mathrm{AIC} = 2k - 2\ln L,
\]
where $k$ is the number of free parameters and $L$ is the maximum likelihood. Models with smaller AIC values are preferred. The criterion penalizes models with more parameters to reduce overfitting, but does not assume that the true model is among those being tested. For two models one compares the difference $\Delta\mathrm{AIC}=\mathrm{AIC}_i-\mathrm{AIC}_j$; on a commonly used (Jeffreys-type) scale, $|\Delta\mathrm{AIC}|$ in the range $2$--$6$ is read as positive evidence for the favored model, $6$--$10$ as strong evidence, and $>10$ as very strong evidence. We stress that this interpretation presupposes that the AIC is computed from a proper likelihood; when AIC is instead evaluated for a surrogate regression (as in the main text), $\Delta\mathrm{AIC}$ measures relative surrogate fit quality and does not by itself carry this likelihood-based evidential interpretation. 
\item[\textbf{BIC (Bayesian information criterion)}:] 
A model selection criterion similar to AIC but with a stronger penalty for model complexity. It is defined as
\[
\mathrm{BIC} = k\ln n - 2\ln L,
\]
where $k$ is the number of parameters, $n$ is the number of data points, and $L$ is the maximum likelihood. Lower BIC values indicate a preferred model. Compared with AIC, BIC more strongly favors simpler models when the dataset is large.

\end{description}

\newpage

\end{document}